\documentclass[useAMS,usenatbib,a4paper]{mnras}

\usepackage{epsfig,bm}
\usepackage{amsbsy,amssymb,amsmath}
\usepackage{graphicx}
\usepackage{dcolumn}

\usepackage{float}
\usepackage[usenames,dvipsnames]{xcolor}
\usepackage{hyperref}
\hypersetup{
    colorlinks = true,
    citecolor = {MidnightBlue},
    linkcolor = {BrickRed},
    urlcolor = {BrickRed}
}

\newcommand{\nv}{\hat{\bf n}}

\newcommand{\fnl}{f_{\rm NL}}
\newcommand{\est}{\epsilon}
\newcommand{\auto}{\est_{\rm A}}
\newcommand{\cross}{\est_{\rm X}}
\newcommand{\estboth}{\est_{\rm A,~X}}
\newcommand{\CV}{\est_{{\rm CV}{}^+}}
\newcommand{\opt}{\est_{\rm opt}}
\newcommand{\autol}{\est_{{\rm A},\ell}}
\newcommand{\crossl}{\est_{{\rm X},\ell}}

\title[Simulated multi-tracer analyses with HI intensity mapping]
{Simulated multi-tracer analyses with HI intensity mapping}
\author[A. Witzemann et al.]{
A. Witzemann$^1$\thanks{E-mail: }, D. Alonso$^{2,3}$, J. Fonseca$^1$, M.G. Santos$^{1,4}$
\\
$^{1}$Department of Physics and Astronomy, University of the Western Cape, Cape Town, 7535, South Africa\\
$^{2}$School of Physics and Astronomy, Cardiff University, The Parade, Cardiff, CF24 3AA, UK\\
$^{3}$University of Oxford, Denys Wilkinson Building, Keble Road, Oxford OX1 3RH, UK\\
$^{4}$SKA SA, The Park, Park Road, Pinelands 7405, South Africa\\
}
\begin{document}

\maketitle

\begin{abstract}
We use full sky simulations, including the effects of foreground contamination and removal, to explore multi-tracer synergies between a SKA-like 21cm intensity mapping survey and a LSST-like photometric galaxy redshift survey. In particular we study ratios of auto and cross-correlations between the two tracers as estimators of the ratio of their biases, a quantity that should benefit considerably from the cosmic variance cancellation of the multi-tracer approach. We show how well we should be able to measure the bias ratio on very large scales (down to $\ell \sim 3$), which is crucial to measure primordial non-Gaussianity and general relativistic effects on large scale structure. We find that, in the absence of foregrounds but with realistic noise levels of such surveys, the multi-tracer estimators are able to improve on the sensitivity of a cosmic-variance contaminated measurement by a factor of $2-4$. When foregrounds are included, estimators using the 21cm auto-correlation become biased. However, we show that cross-correlation estimators are immune to this and do not incur in any significant penalty in terms of sensitivity from discarding the auto-correlation data. However, the loss of long-wavelength radial modes caused by foreground removal in combination with the low redshift resolution of photometric surveys, reduces the sensitivity of the multi-tracer estimator, albeit still better than the cosmic variance contaminated scenario even in the noise free case. Finally we explore different alternative avenues to avoid this problem.
\end{abstract}


\section{Introduction}\label{sec:intro}
  Probing the physics of the primeval Universe is one of the main drivers for observational studies of the cosmos. The Gaussianity of the primordial cosmological perturbations remains an open question which provides further insight into the details of the dynamics of the very early Universe. The current state of the art are the Planck bounds derived from the Cosmic Microwave Background (CMB) \citep{2016A&A...594A..13P}. As an example, the bounds on local-type primordial non-Gaussianity (PNG) yield $\fnl\simeq0.8\pm 5.0$. Furthermore, local PNG introduces a scale dependence in the bias between the Dark Matter (DM) halos and the astrophysical objects used as tracers of the matter distribution \citep{PhysRevD.77.123514,Matarrese:2008nc}.  

  This scale dependence becomes relevant on large cosmological (horizon) scales. At the same time, general-relativistic effects become important on such ultra-large scales (past the matter-radiation equality peak), opening the possibility of testing the theory of gravity in this new regime and find possible hints of deviations to General Relativity (for a comprehensive review on "GR effects" see e.g. \cite{2011PhRvD..84d3516C,2011PhRvD..84f3505B,2014CQGra..31w4002B}). The next generations of Large Scale Structure (LSS) surveys such as the Square Kilometer Array (SKA)\footnote{www.skatelescope.org}, Euclid\footnote{www.euclid-ec.org} and the Large Synoptic Survey Telescope (LSST)\footnote{www.lsst.org}, promise to be able to target such effects by observing ever larger volumes of the Universe. Indeed, the forecasts for the next-generation surveys will improve on the Planck constraint on PNG \citep[see, e.g.][]{Giannantonio:2011ya,Camera:2013kpa,Camera:2014bwa,Alonso:2015uua,Raccanelli:2015vla}. Despite the improvements, forecast errors on local PNG from single tracers of the matter distribution will still be unable to push $\sigma(\fnl)$ below (or close to) 1, crucial to distinguishing between single-field and multi-field inflation \citep[see, e.g.][]{dePutter:2016trg}.

  The crucial limitation on these surveys comes from cosmic variance, which limits measurements on the largest scales. A decade ago \citet{2009PhRvL.102b1302S} proposed a statistical method, often referred to as the multi-tracer technique, to overcome cosmic variance (see also \citealt{McDonald:2008sh, Hamaus:2011dq, Abramo:2013awa}). The basic idea is that, if we only care about effects on the bias of the dark matter tracers and not on dark matter itself, then, by comparing two tracers, we can at least measure the ratio of their bias without requiring to measure the underlying dark matter distribution they trace. This will then avoid cosmic variance, caused by the stochasticity in the particular realization of the matter distribution we observe. By cancelling cosmic variance, we also shift the target set-up of future surveys to probe these large scale effects, since smaller volumes with low noise (e.g. large integration times or higher number densities) are preferred as opposed to huge volumes that sample the modes of interest many times (as long as such smaller volumes include the target scales). 

  Several authors have extensively used the technique to forecast how combinations of future surveys and different DM tracers will impact on the prospects of measuring $\fnl$ as well as other horizon-scale GR effects \citep{PhysRevD.86.063514,Ferramacho:2014pua, Yamauchi:2014ioa,PhysRevD.92.063525,2015ApJ...812L..22F,Fonseca:2016xvi,Abramo:2017xnp,Fonseca:2018hsu,2018PhRvD..97l3540S}. While some combinations do not break the $\sigma(\fnl)<1$ threshold, others have the potential to provide transformational constraints on $\fnl$ and GR effects. Such technique thus opens a new window to probe the physics of inflation and General Relativity with near-future experiments.

  Despite the plethora of works studying the potential and applicability of the multi-tracer technique, little has been done to test and assess the performance of the technique within realistic observational settings for future surveys (although the technique has been employed in some analysis of current data \citep{2013MNRAS.436.3089B,2014MNRAS.437.1109R,2016MNRAS.455.4046M}, with an emphasis on redshift-space distortions). Questions on what estimators to use and whether they will be biased by contaminants still remain unanswered. This paper attempts to address some of these technical and practical issues. We will focus on the combination of an HI intensity mapping (IM) survey carried out by a SKA-like facility \citep{2015aska.confE..19S} with a LSST-like photometric galaxy survey \citep{2009arXiv0912.0201L}. This is a natural combination choice since both surveys will observe the largest cosmological volumes in an overlapping region of the sky in both the radio and optical/infra-red regimes. Moreover, such surveys will be affected by different sky systematics. Intensity mapping of the 21cm emission line of neutral Hydrogen is contaminated by signal from galactic synchrotron emission, free-free emission from galactic and extra-galactic origin and point sources. \citet{doi:10.1093/mnras/stu1666} compiled all the potential radio foregrounds and tested methods to subtract such contaminants from the HI temperature fluctuations. On the other hand, optical galaxy surveys will be affected by galactic dust extinction and star contamination, as well as several observational systematics, which can affect the observed clustering on large scales \citep{2011MNRAS.417.1350R}. The hope is that a combination of foreground cleaning methods and cross-correlations between surveys can help to make measurements that are reasonably free from such contaminants. Moreover, the specific scale-dependence of the cosmological effects might be used to disentangle this signature from any contaminant residuals.

  In this paper we explore the multi-tracer technique in the presence of foregrounds in the HI intensity maps using realistic simulations of the observational process. For this purpose we will construct estimators of the bias ratios and assess their performance at each redshift bin. For simplicity we will neglect the presence of PNG on the tracer biases, making the bias ratios scale independent. Note that PNG should have a negligible impact on the extracted estimator errors. We will focus on IM foregrounds, which are likely to be the main contaminant, and neglect the effects of possible systematics in the optical data for now. Crucially, we wish to determine how sensitive the cancellation of cosmic variance is to IM foreground cleaning and to the observational specifications of each experiment.

  The paper is organized as follows: in section \ref{sec:theory} we discuss possible multi-tracer estimators that can be used to extract the bias ratio of the two tracers and in particular focus on estimators that can be free from foreground or systematic contamination. In section \ref{sec:method} we describe the simulations done for both experiments (SKA1-MID and LSST) and the foreground cleaning method. In section \ref{sec:res} we discuss the results, addressing the performance and errors on the estimators and possible biases. In particular, we discuss the limitations of the current approach and show possible avenues to improve on this technique. We conclude in section \ref{sec:discussion}.

\section{Multi-tracer estimators}\label{sec:theory}
  \subsection{Signal modelling}\label{ssec:theory.signal}
    Our basic observable is the projected fluctuation of a given tracer of the matter distribution on the sky $\Delta(\nv)$. Under the assumption that, on sufficiently large scales, $\Delta(\nv)$ is linearly related to the matter overdensity $\delta_M(t,{\bf x})$, the relation between both quantities can be modelled as:
    \begin{equation}
      \Delta(\nv)=\int_0^\infty dz\,b(z)\,\phi(z)\,\delta_M(t(z),\chi(z)\nv),
    \end{equation}
    where $\chi$ is the comoving radial distance, and $b(z)$ and $\phi(z)$ are the bias and selection functions associated with this tracer. For simplicity we have neglected the contributions from redshift-space distortions, magnification and other relativistic effects \citep{2011PhRvD..84d3516C,2011PhRvD..84f3505B,PhysRevD.85.041301,PhysRevD.85.023504, PhysRevD.86.063514, PhysRevD.87.064026, PhysRevD.88.023502}. This simplifying approximation should not have any significant impact on the final results presented here, since RSDs are suppressed by the broad redshift kernels used and all other terms are highly sub-dominant \citep{ doi:10.1093/mnras/stu2491,doi:10.1093/mnras/stu2474,PhysRevD.92.063525,2015ApJ...812L..22F}.
    
    Given two tracers $a$ and $b$, the angular cross-power spectrum is defined as the two-point function of their harmonic coefficients, and can be related to the matter power spectrum $P(k,z)$ as \citep{2013JCAP...11..044D}:
    \begin{align}
      C^{ab}_\ell\equiv\langle\Delta_{\ell m}^a\Delta_{\ell m}^{b*}\rangle=\frac{2}{\pi}\int_0^\infty dk\,k^2W^a_\ell(k)W^b_\ell(k),\\
      W^a_\ell(k)=\int dz\,b_a(z)\phi_a(z)j_\ell(k\chi(z))\sqrt{P(k,z)},
    \end{align}
    Under Limber's approximation \citep{1954ApJ...119..655L,2008PhRvD..78l3506L}, this expression can be simplified to
    \begin{align}
      C^{ab}_\ell=\int d\chi \frac{b_a\phi_a\,b_b\phi_b\,H^2(\chi)}{\chi^2}\,P\left(z(\chi),k=\frac{\ell+1/2}{\chi}\right),
    \end{align}
    where $H$ is the expansion rate.
    
    In this analysis we have used two different types of tracers: the overdensity of galaxy number counts, which we will label as $\Delta^g$, and the temperature fluctuations in the 21cm line emission caused by neutral hydrogen (HI), $\Delta^{\rm H}$. In the case of galaxy clustering, we approximate the linear galaxy bias as $b^{\rm g} = 1 + 0.84z$ \citep{2009arXiv0912.0201L}, which is an estimate of the results from \citep{0004-637X-601-1-1}. On the other hand, as described in Section \ref{ssec:method.fgs}, the presence of spectrally smooth radio foregrounds makes it infeasible to measure the average 21cm brightness temperature $\bar{T}_{21}$, and it is therefore completely degenerate with the linear bias function associated with this tracer: $b_{\rm H}(z)=\bar{T}_{21}(z)\,b_{\rm HI}(z)$, where $b_{\rm HI}$ is the linear clustering bias associated with the cosmic overdensity of neutral hydrogen \footnote{Note that although this is the case in our analysis, there do exist ways to extract the average HI brightness temperature, for example using cross-correlations, or HI galaxy surveys \citep{doi:10.1093/mnras/stx1388}.}. We model both quantities after \cite{0004-637X-803-1-21}.

    Finally, the observed fluctuations $\Delta^a$ are inevitably contaminated by noise. In the case of galaxy clustering, this is associated with shot-noise due to the discrete nature of the sources used to reconstruct the true underlying distribution. In this case, the noise power spectrum is simply given by the inverse number density of tracer sources in units of Sr$^{-1}$,
    \begin{equation}
      N_\ell^{gg}=\frac{1}{\bar{n}}.
    \end{equation}
    For 21cm, the combination of instrumental noise and beam smoothing, caused by the telescope's finite size, effectively erases all modes below the telescope resolution. For an angular Gaussian beam, the harmonic coefficients of the beam, multiplying the signal in harmonic space, can be simply modelled as
    \begin{equation}
      B_\ell=\exp\left(-\frac{\ell(\ell+1)\theta_{\rm FWHM}^2}{16\log2}\right),
    \end{equation}
    where $\theta_{\rm FWHM}$ if the beam full-width at half-maximum (FWHM) at a given frequency. The instrumental noise can then be modelled as an additive Gaussian random field with flat power spectrum. For single-dish observations, this is simply given by \citep{0004-637X-803-1-21}
    \begin{equation}
      N^{\rm HH}_\ell=\frac{T_{\rm sys}^24\pi f_{\rm sky}}{N_{\rm dish}\Delta\nu t_{\rm tot}}.
    \end{equation}
    Here $T_{\rm sys}$ is the system temperature, $f_{\rm sky}$ is the total observed sky fraction, $N_{\rm dish}$ is the number of dishes in the instrument, $t_{\rm tot}$ is the total integration time and $\Delta\nu$ is the frequency bandwidth for the particular sky map under consideration.

    It is worth noting that we assume no cross-noise term between galaxies and HI. This is expected to be present if the HI-emitting star-forming galaxies form a significant fraction of the galaxy sample, however we assume this shot-noise contribution to be subdominant. We also neglect any correlated $1/f$-like noise component for intensity mapping. We refer the reader to \cite{2018MNRAS.478.2416H} for a more detailed discussion of correlated noise in the context of foreground contamination and removal.

  \subsection{The Surveys}\label{ssec:theory.surveys}
    \begin{figure}
      \includegraphics[width=1.0\columnwidth]{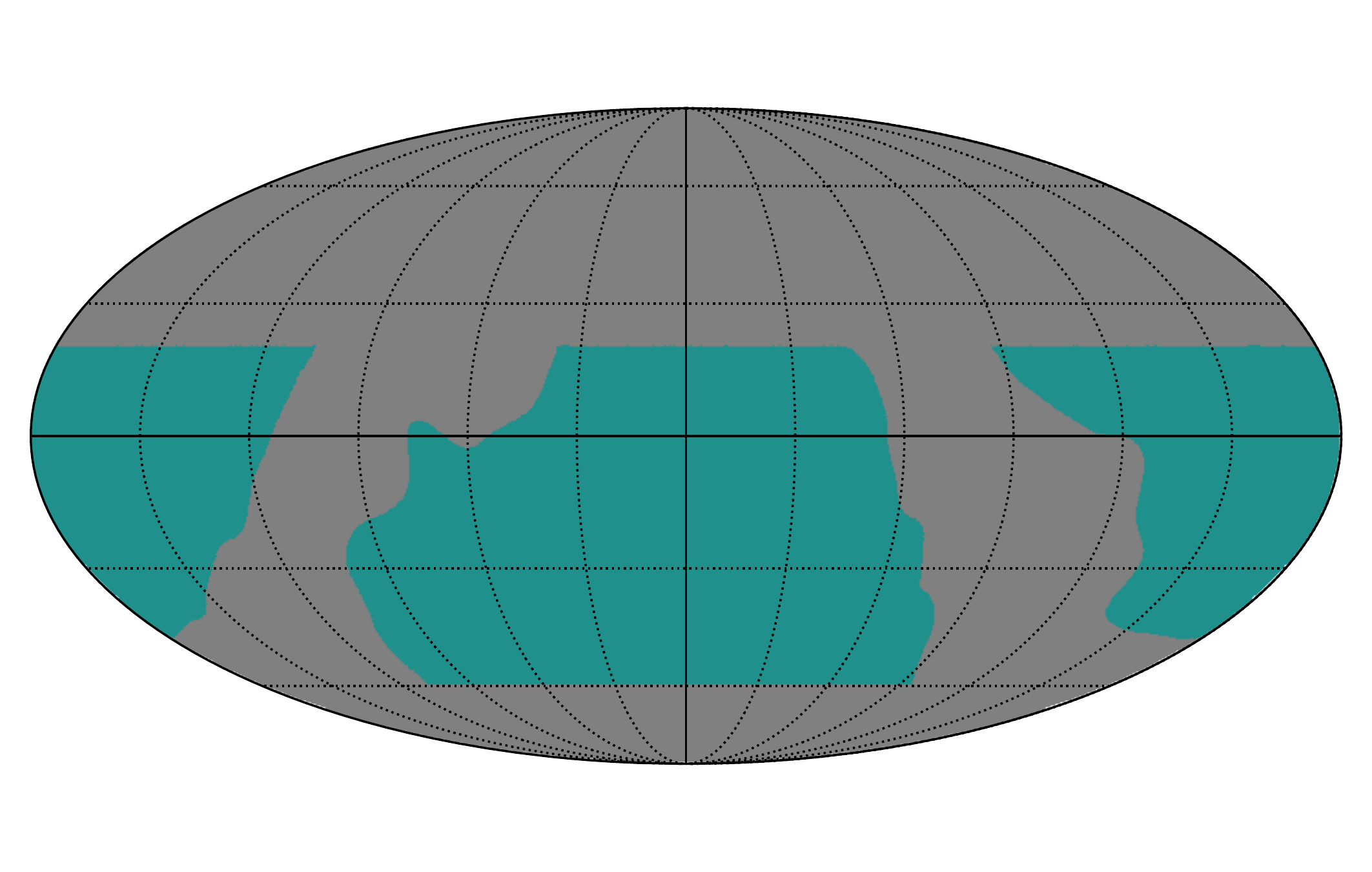}
      \caption{Sky mask used in our analysis, shown in Mollweide's projection and equatorial coordinates. The masked area is shown in grey. The footprint corresponds to the sky observable from the LSST and SKA with the regions of highest galactic emission (both in synchrotron and dust) removed. The total unmasked area is $16900 {\rm ~deg}^2$ ($f_{\rm sky}=0.41$.) }
    \label{fig:mask}
    \end{figure}    
    Our forecasts focus on the combination of 21cm intensity maps, constructed from the SKA data, with optical observations of the galaxy distribution as could be achieved by LSST. We describe the models used for both datasets here.
    
    We assume the first phase of SKA (in particular SKA-1 MID \citep{2015aska.confE..19S}) to consist of 197 dishes, which will use a total of $t_{\rm tot}=10.000$ h integration time to produce intensity maps covering $\sim60\%$ of the sky. We assume a combination of surveys carried out with band 1 and band 2 receivers, and we use a frequency range of $\nu\in(390,1300)\,{\rm MHz}$, corresponding to a redshift interval $0.1\leq z\leq2.65$. Since we work with individual redshift bins at a time, our results are always valid for the receiver type that covers the relevant redshift range. We will assume single-dish observations, which are limited in angular resolution by a beam that we model as Gaussian with a FWHM given by $\theta_{\rm FWHM}=1.22\,\lambda/D_{\rm dish}$, where $\lambda$ is the observed wavelength and $D_{\rm dish}$ is the dish diameter. We assume a diameter $D_{\rm dish}=14.5$m\footnote{SKA-1 MID will consist of a combination of $15$ and $13.5$m dishes, and we use $14.5$ as an approximation to the mean dish diameter. This choice should not affect the final results of this study.}. Finally, we add white noise as described in the previous section, with a smoothly-varying system temperature $T_{\rm sys}$ following the values given in \cite{2017arXiv170906099S}. Further particulars regarding the specific simulated intensity maps used in this analysis are described in Section \ref{ssec:method.sims}

    For LSST, we use the redshift distribution modelled in \cite{doi:10.1093/mnras/stu2474}, which yields an integrated number density of $43$ galaxies per arcmin$^2$, in agreement with \cite{2009arXiv0912.0201L}. As described in Section \ref{ssec:method.sims}, we do not make a precise modelling of the photometric redshift accuracy that LSST will achieve, and instead work with redshift bins wide enough ($\Delta z=0.1$) to simulate the loss of small radial scales. We do this in order to facilitate the interpretation of the auto-correlation and cross-correlation estimators presented in the next section. A more realistic treatment would either account for the difference in radial window function between the 21cm and optical bins, or re-weight the 21cm frequency channels contributing to each bin to mimic the photo-$z$ window function as closely as possible.

    We assume almost complete overlap between SKA and LSST, given their common observable sky. After accounting for contamination from galactic synchrotron (radio) and dust (optical), the final common footprint, displayed in Fig. \ref{fig:mask}, covers $41\%$ of the sky.

  \subsection{The Estimators}\label{ssec:theory.estimators}
    Under the assumption that the bias functions vary slowly over the support of the selection functions, and in the limit where the selection functions for both tracers are the same ($\phi^g=\phi^{\rm H}\equiv\phi$), the three different auto and cross-power spectra described in Section \ref{ssec:theory.signal} can be written as:
    \begin{align}\nonumber
      &C^{gg}_\ell=b_g^2\,C_\ell+N^{gg}_\ell,\\
      &C^{{\rm H}g}=b_g\bar{T}_{21} b_{\rm HI}\,B_\ell\,C_\ell,\label{eq:cls}\\\nonumber
      &C^{\rm HH}_\ell=\bar{T}^2_{21}b^2_{\rm HI}\,B_\ell^2\,C_\ell+N^{\rm HH}_\ell,
    \end{align}
    where $C_\ell$ is the angular power spectrum of the matter overdensity projected along the line of sight with $\phi$.
    
    On a realization-by-realization basis, the measured values of these quantities will be subject to sample variance, due to the stochastic nature of both the underlying matter fluctuations and the instrumental and shot noise. For signal-dominated modes, the realization-dependent fluctuations will coincide for the three power spectra, and therefore it is possible to constrain certain parameters beyond the limit imposed by sample variance if only a single tracer was available \citep{2009PhRvL.102b1302S}. One obvious example of this is the ratio of the tracer bias functions, which in an ideal noiseless case could be measured exactly by taking ratios of the power spectra above. In this work we will focus on the quantity
    \begin{equation}\label{eq:defeps}
      \est\equiv\frac{b_{\rm HI}\bar{T}_{21}}{b_g},
    \end{equation}
    for which we propose two different estimators:
    \begin{align}\label{eq:est_e1}
      &\hat{\est}_{{\rm A},\ell}\equiv\sqrt{\frac{\hat{C}_\ell^{\rm HH}-N^{\rm HH}_\ell}{B_\ell^2\left(\hat{C}^{gg}_\ell-N^{gg}_\ell\right)}},\\\label{eq:est_e2}
      &\hat{\est}_{{\rm X},\ell}\equiv      \frac{\hat{C}_\ell^{{\rm H}g}}{B_\ell\left[\hat{C}^{gg}_\ell-N^{gg}_\ell\right]},
    \end{align}
    where all hatted quantities (e.g. $\hat{C}^{\rm HH}_\ell$) are measurements in a given realization. In addition to this, we will also consider a third estimator making use of both the auto and cross-correlation, which combines $\estboth$ in an inverse-variance-weighted manner:
    \begin{equation}
      \opt=\frac{\sum_{i,j}{\sf C}^{-1}_{ij}\est_j}{\sum_{ij}{\sf C}^{-1}_{ij}},
      \label{eq:est_opt}
    \end{equation}
    where ${\sf C}$ is the covariance matrix of the two previous estimators computed from simulations.

    These three estimators can be understood as different limits of a more general maximum-likelihood estimator combining the three cross-correlations simultaneously, which allow us to explore the impact of foreground contamination in the 21cm maps. 

\section{Simulated Forecasts}\label{sec:method}
  \subsection{The Simulations}\label{ssec:method.sims}
    \begin{figure*}
      \includegraphics[width=1.0\textwidth]{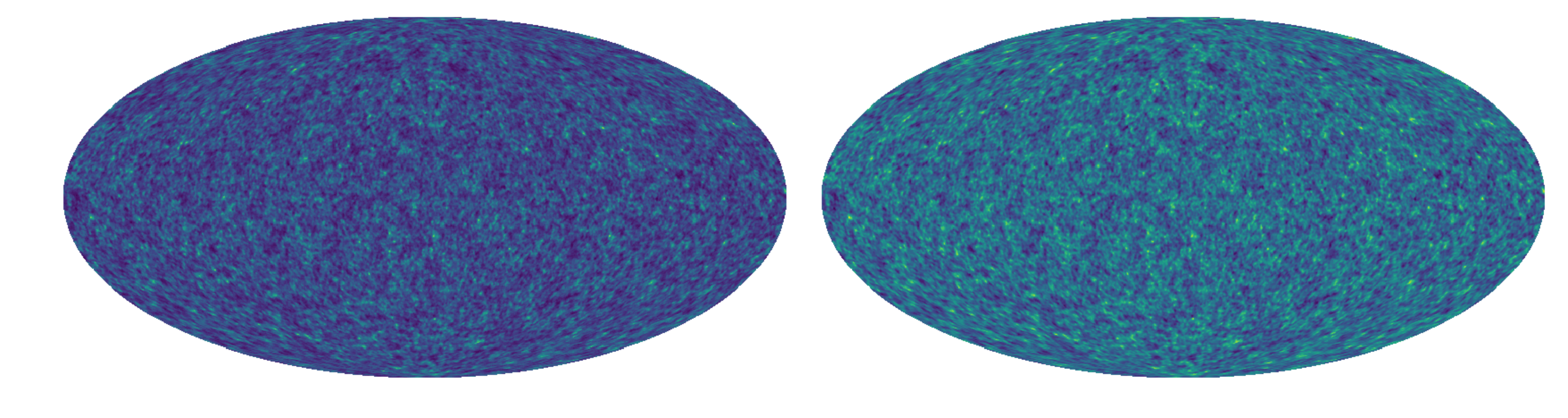}
      \caption{Galaxy (left) and HI map at $1050$ MHz ($z\simeq0.35$) with a redshift bin width of $\Delta z = 0.1$. We choose a low-redshift bin in order to make the tight correlation between both maps more visually apparent. Both maps trace the same DM background, and show the pure cosmological signal, before inclusion of foregrounds, beam smoothing or noise. Note that the HI maps were generated in much thinner bins of $\Delta \nu = 1$ MHz. Noise, beam and foreground simulation was done in these thin bins, and those maps were later merged to match the thicker bins of the galaxy maps.}
      \label{fig:maps}
    \end{figure*}
    We produce synthetic signal simulations of both the galaxy distribution and 21cm maps using the publicly available code {\tt CoLoRe}\footnote{\url{https://github.com/damonge/CoLoRe}}. {\tt CoLoRe} efficiently generates intensity maps for any arbitrary line-emitting species and source catalogues tracing the same dark matter distribution (with their respective biases $b_{\rm HI}(z)$ and $b_{\rm gal}(z)$). {\tt CoLoRe} first generates a Gaussian realization of the linearised density field at $z=0$ along with the corresponding linear radial velocity field. It then linearly evolves density and velocity to the redshift of each grid point in the simulation and produces a 3D cube of the physical matter density in the lightcone using a log-normal transformation (described in e.g. \cite{1991MNRAS.248....1C})\footnote{Note that {\tt CoLoRe} is also able to produce physical density fields through other more accurate methods (e.g. 1$^{\rm st}$ and 2$^{\rm nd}$-order Lagrangian perturbation theory, but we chose the log-normal for simplicity and performance reasons). This choice should be irrelevant given that our analysis focuses on relatively large scales.} For the galaxy sample the density field is biased and then Poisson-sampled using the galaxy number density $N(z)$. For 21cm, the density field is used to generate a biased HI density, which we then interpolate into spherical shells that we output as sky maps. For simplicity, we switch off the effect of redshift-space distortions, and therefore the redshift of each source is calculated without accounting for the local velocity field. We simulate a cubic box with $2048^3$ Cartesian grid points and a length large enough to encompass the comoving volume to redshift $z=2.7$. This yields a grid resolution of $\Delta x\simeq4\,h^{-1}{\rm Mpc}$.  The initial Gaussian density field is smoothed with a Gaussian kernel of size $R_G=5\,h^{-1}{\rm Mpc}$ to avoid grid artifacts as well as the non-linear distortions induced by the log-normal transformation. This scale is significantly smaller than those we focus on, or than the SKA beam, and therefore the impact of this smoothing on our results is negligible.

    We generate 21cm intensity maps with a frequency resolution of $\Delta\nu = 1$ MHz. To each of these maps we first add the simulated foreground maps, smooth them using the SKA Gaussian beam and add the instrumental noise as described above. To study the case of ideal noise-free cosmic-variance cancellation we also simulate equivalent maps of the galaxy overdensity without shot noise. We simulate these as an alternative intensity mapping species with unit mean temperature and a bias given by the galaxy bias. The foregrounds are simulated using {\tt ForGet}, part of the publicly available {\tt CRIME} package\footnote{\url{https://github.com/damonge/CRIME}} \citep{doi:10.1093/mnras/stu1666}. We consider 4 unpolarized foreground sources, including galactic synchrotron, galactic and extragalactic free-free emission and extragalactic point sources.

    From these outputs we produce maps of the 21cm temperature fluctuations and of the galaxy overdensity on thin radial bins with an equivalent frequency width $\Delta\nu=1\,{\rm MHz}$. After the foreground cleaning stage, described in section \ref{ssec:method.fgs}, the resulting 21cm maps are merged to thicker bins with a width of $\Delta z = 0.1$, and the same is done to estimate the galaxy overdensity in bins of the same width.

    Finally, in order to study the statistical properties of our estimators, we generate  $N_{\rm sim} = 200$ simulations of the dark matter background, using different seeds for the Gaussian density field. Each simulation is populated with the HI and galaxy distributions, using different seeds for the noise realization and foreground maps. All simulations assume a $\Lambda$CDM cosmological model with parameters $(\Omega_M,\Omega_b,n_s,\sigma_8,h)=(0.3,0.05,0.96,0.8,0.7)$. 
   
    Figure \ref{fig:maps} shows simulated maps of the galaxy overdensity (left) and the HI temperature (right) using this procedure at a redshift $z\simeq0.35$. Both maps are very strongly correlated, and display similar structures. This tight correlation is the basis for the cosmic-variance cancellation implicit in multi-tracer studies.
    
  \subsection{Foreground Removal}\label{ssec:method.fgs}
    Foreground removal methods for 21cm intensity mapping \citep{2013MNRAS.429..165C,2014MNRAS.441.3271W,2014ApJ...781...57S,2015PhRvD..91h3514S,doi:10.1093/mnras/stu2474,2018arXiv180104082Z} try to separate the cosmological and foreground signals by making use of their different spectral properties: while foregrounds are expected to have a smooth dependence with frequency, which should also be highly correlated across the sky, the cosmological signal follows the large-scale structure, and therefore contains power across a large range of Fourier scales (both in frequency and angles).
    
    Let ${\bf d}$ be a vector containing our measurements of the brightness temperature along a fixed line of sight. In general it will contain contributions from foregrounds ${\bf f}$, cosmological signal ${\bf c}$ and instrument noise ${\bf n}$:
    \begin{equation}
      {\bf d}={\bf f}+{\bf c}+{\bf n}={\bf f}+{\bf s},
    \end{equation}
    where we have grouped all noise-like components into ${\bf s}\equiv{\bf c}+{\bf n}$. Most foreground removal methods recover an estimate of ${\bf s}$ by linearly filtering the data:
    \begin{equation}
      {\bf s}_c={\sf W}\cdot{\bf d},
    \end{equation}
    using a filter ${\sf W}$ that minimizes the presence of foreground residuals on ${\bf s}$. For instance, principal component analysis (PCA) corresponds to a filter ${\sf W}=1-{\bf U}_{\rm PC}$, where ${\sf U}_{\rm PC}$ is the matrix of principal eigenvectors of the data covariance matrix. As another example, a linear fit to a set of smooth functions of frequency, stored in the columns of a matrix ${\sf A}$, would correspond to a choice of filter
    \begin{equation}
      {\sf W}={\sf 1}-{\sf A}\left({\sf A}^T{\sf S}{\sf A}\right)^{-1}{\sf A}^T{\sf S}^{-1},
    \end{equation}
    where ${\sf S}$ is the covariance of ${\bf s}$.

    After filtering, the cleaned signal
    \begin{equation}\label{eq:post_fgr}
      {\bf s}_c={\sf W}{\bf s}+{\sf W}{\bf f}
    \end{equation}
    will contain both a version of the original signal where typically the longer-wavelength radial modes have been downweighted (${\sf W}{\bf s}$), as well as foreground residuals (${\sf W}{\bf f}$), unless a perfect knowledge of the foreground spectral behaviour can be achieved. This has two main consequences when it comes to using ${\sf s}_c$ for cosmology:
    \begin{itemize}
      \item Unless foregrounds have been perfectly removed (which is never the case), the auto-correlation of the 21cm data will be contaminated by foreground residuals that must be marginalized over (unless we can convince ourselves that their amplitude lies below the noise level at the relevant length scales).
      \item Even when cross-correlating with other tracers of the large-scale structure, the loss of radial modes implied by the filter ${\sf W}$ must be taken into account and corrected for in the model for the cross-correlation.
    \end{itemize}    
    The first effect is inherent to 21cm auto-correlations, and can only be overcome if the residual contamination is sufficiently small, or if a sufficiently accurate foreground model can be built to marginalize over their contribution. However, since we always know the filter ${\sf W}$ used by the foreground cleaning pipeline, the second effect can be modelled and taken into account. In general, the action of ${\sf W}$ will be to remove power from the largest radial scales, thus reducing the overall amplitude of any projected clustering statistic. Characterizing this reduction exactly requires a full model of the 3D power spectrum, however we will take a simpler approximate method here, similar to the procedure used in e.g. \cite{2013ApJ...763L..20M,2013MNRAS.434L..46S}. We model the impact of ${\sf W}$ on the angular power spectrum as a scale-dependent, multiplicative transfer function $T_\ell$. I.e.:
    \begin{equation}
      \tilde{C}^{\rm HH}_\ell=T_\ell^2\,C^{\rm HH}_\ell,\hspace{12pt} \tilde{C}^{\rm Hg}_\ell=T_\ell\,C^{{\rm H}g}_\ell.
      \label{eq:transfer}
    \end{equation}
    Here $\tilde{C}_\ell$ and $C_\ell$ denote power spectra computed after foreground removal and in the absence of foregrounds respectively, and $\tilde{C}^{\rm HH}_\ell$ does not include the contribution from foreground residuals (i.e. it is only the auto-correlation of the first term in Eq. \ref{eq:post_fgr}). We estimate the transfer function from our simulations as:
    \begin{equation}
      T_\ell=\frac{\langle C^{\tilde{\rm H}{\rm H}}_\ell\rangle-N^{\tilde{\rm H}{\rm H}}_\ell}{\langle C^{\rm HH}_\ell\rangle-N^{{\rm HH}}_\ell},
      \label{eq:filter}
    \end{equation}
    where $C^{\rm HH}_\ell$ is the auto-correlation of a foreground-free simulation, $C^{\tilde{\rm H}{\rm H}}_\ell$ is the cross-correlation between a foreground-cleaned and a foreground-free simulation (we have subtracted the noise bias from both power spectra), and $\langle\,\rangle$ denotes averaging over all simulations.
    
    After accounting for this loss of modes, the estimators $\estboth$ in Equations \ref{eq:est_e1} and \ref{eq:est_e2} above become
    \begin{align}\label{eq:est_e1b}
      &\autol\equiv\sqrt{\frac{\hat{C}_\ell^{\rm HH}-N^{\rm HH}_\ell}{(T_\ell B_\ell)^2\left(\hat{C}^{gg}_\ell-N^{gg}_\ell\right)}},\\\label{eq:est_e2b}
      &\crossl\equiv      \frac{\hat{C}_\ell^{{\rm H}g}}{T_\ell\,B_\ell\left[\hat{C}^{gg}_\ell-N^{gg}_\ell\right]},
    \end{align}

\section{Results}\label{sec:res}
  \subsection{Theoretical expectation}\label{ssec:res.theory}
    Before we set off to use our simulations to study the feasibility of multi-tracer methods for intensity mapping, it is instructive to produce a theoretical estimate of the expected performance of our estimators, in order to better understand the simulated results, as well as the main sources of cosmic variance cancellation.
    
    From the expressions for $\auto$ and $\cross$ in Eqs. \ref{eq:est_e1} and \ref{eq:est_e2}, we can write, for one particular realization:
    \begin{align}
      &\hat{\est}_{{\rm A},\ell}=\autol\sqrt{\frac{1+\Delta \hat{C}_\ell^{\rm HH}/(C_\ell^{\rm HH}-N_\ell^{\rm HH})}{1+\Delta \hat{C}_\ell^{gg}/(C_\ell^{\rm gg}-N_\ell^{gg})}},\\
      &\hat{\est}_{{\rm X},\ell}=\crossl\frac{1+\Delta \hat{C}_\ell^{g{\rm H}}/C^{g{\rm H}}_\ell}{1+\Delta \hat{C}_\ell^{gg}/(C_\ell^{gg}-N_\ell^{gg})},
    \end{align}
    where, as before, all hatted quantities (e.g. $\hat{\est}_{{\rm X},\ell}$) are measurements of the equivalent non-hatted observables in a given realization, and $\Delta \hat{C}^{XY}_\ell$ is the fluctuation around the mean $C^{XY}_\ell$ in a given realization. Linearising with respect to these fluctuations, we obtain:
    \begin{align}
      &\frac{\hat{\est}_{{\rm A},\ell}-\autol}{\autol} \approx \frac{1}{2}\left(\frac{\Delta\hat{C}^{\rm HH}_\ell}{C^{\rm HH}_\ell-N^{\rm HH}_\ell}-\frac{\Delta\hat{C}^{gg}_\ell}{C^{gg}_\ell-N^{gg}_\ell}\right),\\
      &\frac{\hat{\est}_{{\rm X},\ell}-\crossl}{\crossl} \approx \frac{\Delta\hat{C}^{g{\rm H}}_\ell}{C^{g{\rm H}}_\ell}-\frac{\Delta\hat{C}^{gg}_\ell}{C^{gg}_\ell-N^{gg}_\ell}.
    \end{align}
    
    To first order, the inverse-squared signal-to-noise ratio can be found by taking the expectation value of the square of the above quantities, obtaining:
    \begin{align}
      \nonumber
      &\left(\frac{S}{N}\right)_{{\rm A},\ell}^{-1}=\frac{1}{2}\left[\frac{{\rm Cov}^{{\rm HH},{\rm HH}}_\ell}{(C^{\rm HH}_\ell-N^{\rm HH}_\ell)^2}+\frac{{\rm Cov}^{gg,gg}_\ell}{(C^{gg}_\ell-N^{gg}_\ell)^2}-\right.\\&\hspace{65pt}\left.2\frac{{\rm Cov}^{{\rm HH},gg}_\ell}{(C^{\rm HH}_\ell-N^{\rm HH}_\ell)(C^{gg}_\ell-N^{gg}_\ell)}\right]^{1/2}\\
      \nonumber
      &\left(\frac{S}{N}\right)_{{\rm X},\ell}^{-1}=\left[\frac{{\rm Cov}^{g{\rm H},g{\rm H}}_\ell}{(C^{g{\rm H}}_\ell)^2}+\frac{{\rm Cov}^{gg,gg}_\ell}{(C^{gg}_\ell-N^{gg}_\ell)^2}-\right.\\&\hspace{55pt}\left.2\frac{{\rm Cov}^{g{\rm H},gg}_\ell}{C^{g{\rm H}}_\ell(C^{gg}_\ell-N^{gg}_\ell)}\right]^{1/2},
    \end{align}
    where ${\rm Cov}^{WX,YZ}_\ell\equiv\langle\Delta\hat{C}^{WX}_\ell\Delta\hat{C}^{YZ}_\ell\rangle$. For Gaussian fields, a simplified estimate of the covariance matrix (that does not account for e.g. survey geometry) is \citep{1995PhRvD..52.4307K}:
    \begin{equation}
      {\rm Cov}^{WX,YZ}_\ell=\frac{C^{WY}_\ell C^{XZ}_\ell+C^{WZ}_\ell C^{XY}_\ell}{(2\ell+1)f_{\rm sky}\Delta\ell},
    \end{equation}
    where $f_{\rm sky}$ is the survey sky fraction and $\Delta\ell$ is the width of the $C_\ell$ bandpowers used in the analysis.
    
    Substituting this result into the equations above we obtain a final expression for the theoretical signal-to-noise ratio:
    \begin{align}
      \nonumber
      &\left(\frac{S}{N}\right)^{-1}_{{\rm A},\ell}=\frac{1}{\sqrt{2n_\ell}}\left[\frac{(C^{\rm HH}_\ell)^2}{(C^{\rm HH}_\ell-N^{\rm HH}_\ell)^2}+\frac{(C^{gg}_\ell)^2}{(C^{gg}_\ell-N^{gg}_\ell)^2}-\right.\\\label{eq:thA}
      &\hspace{80pt}\left.\frac{2(C^{g{\rm H}}_\ell)^2}{(C^{\rm HH}_\ell-N^{\rm HH}_\ell)(C^{gg}_\ell-N^{gg}_\ell)}\right]^{1/2},\\\label{eq:thX}
      &\left(\frac{S}{N}\right)^{-1}_{{\rm X},\ell}=\frac{1}{\sqrt{n_\ell}}\left[1+\frac{C^{\rm HH}_\ell C^{gg}_\ell}{(C^{g{\rm H}}_\ell)^2}-\frac{2C^{gg}_\ell(C^{gg}_\ell-2N^{gg}_\ell)}{(C^{gg}_\ell-N^{gg}_\ell)^2}\right]^{1/2},
    \end{align}
    where $n_\ell\equiv(2\ell+1)f_{\rm sky}\Delta\ell$ is the number of available modes in a given bandpower.
    
    Inspecting Eqs. \ref{eq:thA} and \ref{eq:thX}, the idea of cosmic variance cancellation becomes apparent: for perfectly correlated tracers ($C^{g{\rm H}}_\ell\equiv\sqrt{C^{\rm HH}_\ell C^{gg}_\ell}$), and in the absence of noise ($N^{gg}_\ell,N^{\rm HH}_\ell\rightarrow0$), the negative terms in these equations, originating from the covariance between numerator and denominator in the estimators, exactly cancel the positive terms, and we obtain $(S/N)^{-1}\rightarrow0$. This cosmic variance cancellation would not be possible if the observables entering the estimators were not strongly correlated, as would be the case if, for instance, the 21cm maps and galaxy catalog covered non-overlapping regions of the sky. In this case, ${\rm Cov}^{{\rm HH},gg}_\ell=0$, and the signal-to-noise ratio for a cosmic-variance limited version of $\autol$ would read:
    \begin{equation}\label{eq:thCV}
     \left(\frac{S}{N}\right)^{-1}_{{\rm CV},\ell}=\frac{1}{\sqrt{2n_\ell}}\left[\frac{(C^{\rm HH}_\ell)^2}{(C^{\rm HH}_\ell-N^{\rm HH}_\ell)^2}+\frac{(C^{gg}_\ell)^2}{(C^{gg}_\ell-N^{gg}_\ell)^2}\right]^{1/2}.
    \end{equation}
    
    We will make use of these theoretical estimates (Eqs. \ref{eq:thA}, \ref{eq:thX} and \ref{eq:thCV}) in the next section to validate the results of our simulated results in the absence of foregrounds.

  \subsection{Foreground-free results} \label{ssec:res.nofg}
    \begin{figure}
      \includegraphics[width=1.0\columnwidth]{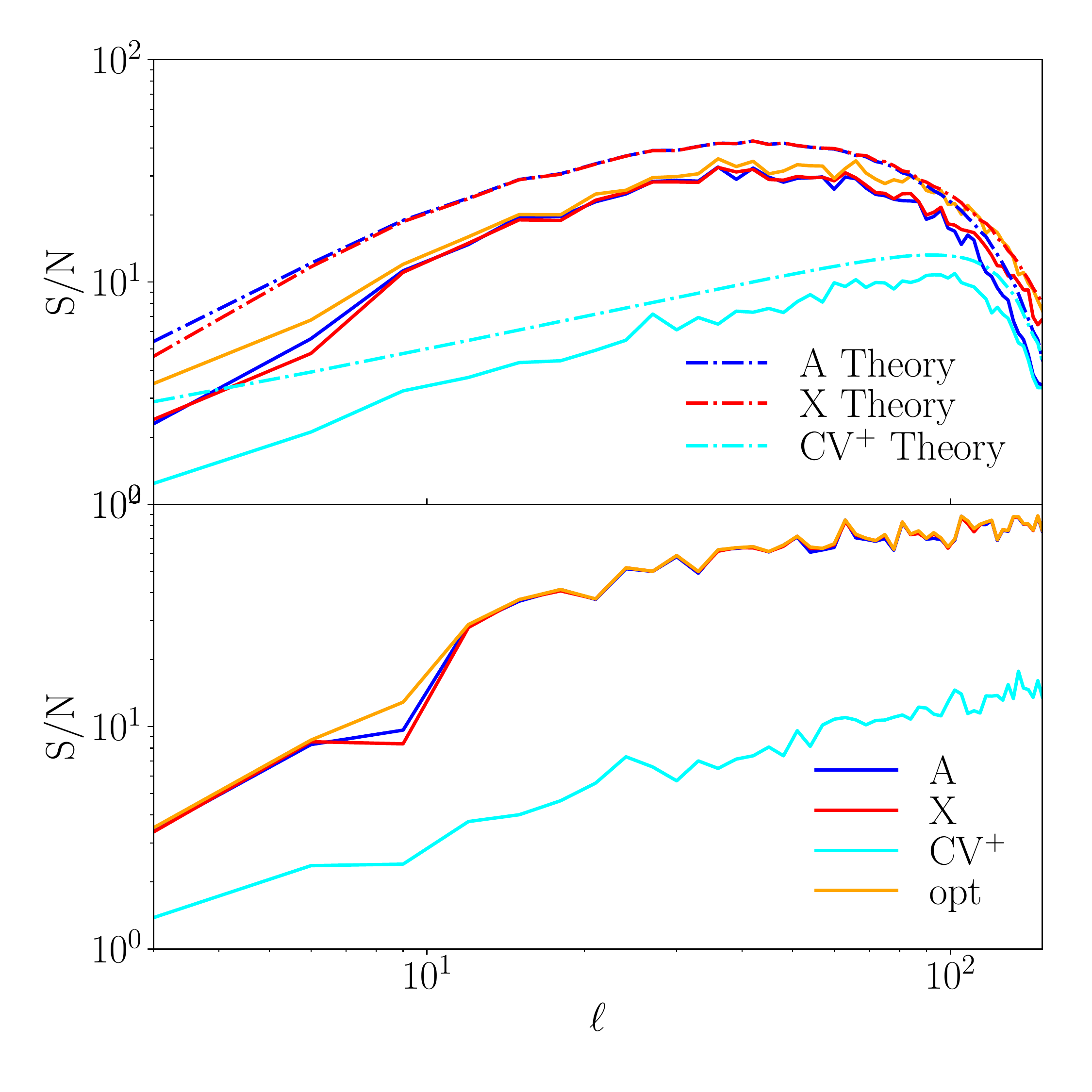}
      \caption{The signal-to-noise ratio for all estimators in the no foregrounds case (top) and no foregrounds, no noise and no beam case (bottom). Results from the simulations (solid lines) give slightly lower signal-to-noise than the theoretical predictions (dotted-dashed lines) in the upper panel from section \ref{ssec:res.theory}. Naturally, $\opt$ (yellow) has the smallest variance, while $\auto$ and $\cross$ perform similarly (blue and red, respectively). All of them beat the cosmic variance estimator $\CV$ (cyan), in the foreground-free case including noise and beam by a factor of 2-4, and in the noiseless case by a factor of 3-8. It is worth noting that little or no sensitivity is lost by discarding all 21cm auto-correlation information and using only cross-correlations (red vs. orange lines).} \label{fig:nofg}
    \end{figure}
    In order to quantify the full power of the cosmic variance cancellation in the estimators described in Section \ref{ssec:theory.estimators}, we first explore the results from simulations without foregrounds or foreground removal, while including noise, masking and beam smoothing. In this case, all the radial modes are present in the HI data (i.e. the transfer function is $T_\ell=1$), and can be used to constrain the bias ratio. The upper panel of Figure \ref{fig:nofg} shows the signal-to-noise ratio of all estimators as a function of multipole $\ell$ for the redshift bin centered around $z=0.8$. For concreteness, the quantity plotted is 
    \begin{equation}
      \left(\frac{\rm S}{\rm N}\right)_\ell=\frac{\est_\ell^{\rm true}}{\sigma_\ell},
    \end{equation}
    where
    \begin{equation}
      \est_{{\rm A},\ell}^{\rm true}\equiv\left.\sqrt{\frac{\langle C_\ell^{\rm HH}\rangle}{\langle C_\ell^{gg}\rangle}}\right|_{\rm FG-free},
      \hspace{6pt}
      \est_{{\rm X},\ell}^{\rm true}\equiv\left.\frac{\langle C_\ell^{\rm Hg}\rangle}{\langle C_\ell^{gg}\rangle}\right|_{\rm FG-free},
    \end{equation}
    and 
    \begin{equation}
      \sigma_\ell^2=\langle \est_\ell^2\rangle-\langle\est_\ell\rangle^2.
    \end{equation}
    Here, angle brackets denote averaging over all simulations. Note that we define $\est^{\rm true}_\ell$ as the value of the estimator found in foreground-free simulations, and not as the bias ratio given in Eq. \ref{eq:defeps}. This is due to the fact that the bias functions and the background 21cm temperature vary slightly within the redshift bin, giving rise to a non-negligible scale dependence of the estimators that would be interpreted as a bias when compared with averages of $\est$ over redshift, even for foreground-free simulations. For comparison, the figure also shows results for an additional estimator $\CV$, defined as a version of $\auto$ in which the auto-power spectra of 21cm and galaxies are computed from simulations with different seeds. The aim of this estimator is to show the results that would be obtained in the absence of cosmic-variance cancellation (e.g. as would be the case when trying to constrain $f_{\rm NL}$ from a single tracer). Note that we calculate $\CV$ in different scenarios, also including instrumental noise, therefore it is not necessarily limited by cosmic-variance.
    
    The signal-to-noise ratio (SNR) of all estimators is shown in the top panel of Figure \ref{fig:nofg}, which shows how it should be possible to significantly increase the sensitivity within the multipole range $\ell\lesssim100$ by a factor of up to $\sim4$ with respect to the CV-dominated case. This is true for both $\auto$ and $\cross$, which achieve very similar sensitivities. The tight correlation between both estimators implies that the improvement associated with combining both into $\opt$ is mild, and that very little information is lost by using only cross-correlation information and discarding the 21cm auto-correlations. For comparison, we show the theoretical predictions derived in the previous section as dashed lines. The theory lines follow the same trends as the simulated results, although they predict a SNR that is $\sim1.3$ times higher than the simulations, owing to the approximations that go into their derivation. In all cases, no significant cosmic variance cancellation can be achieved beyond the scale of the SKA beam ($\ell\sim100$), and the overall SNR drops significantly.
    
    The impact of noise on cosmic-variance cancellation can be further explored in a more idealized scenario, by making use of noiseless maps (i.e. simulations containing no 21cm instrumental noise or galaxy shot noise, as described in Section \ref{ssec:method.sims}). The results, in terms of $S/N$, are shown in the lower panel of Fig. \ref{fig:nofg}. Even in this idealized situation it is not possible to achieve exact cosmic variance cancellation ($S/N\longrightarrow\infty$), and the relative improvement with respect to the CV-dominated case asymptotes at a factor of $\sim4-5$. This is caused by two factors: the redshift evolution of the bias functions within the relatively thick redshift bins, and the non-linear lognormal transformation used by {\tt CoLoRe} to guarantee positive-definite density fields. Both effects produce slight differences in the galaxy and HI maps that prevent exact cosmic variance cancellation. We can only expect the impact of both effects to increase in a more realistic situation, in the presence of uncertain and scale-dependent bias relationships. As expected, the absence of noise allows this level of CV cancellation to be sustained beyond $\ell\sim100$, in comparison with the results described above. 
    
    Although the results presented here are encouraging in terms of the large relative improvement with respect to the CV limit, their validity must be verified when foregrounds are included.
  
  \subsection{Foreground removal} \label{ssec:res.fgrm}
    \begin{figure}
      \includegraphics[width=1.\columnwidth]{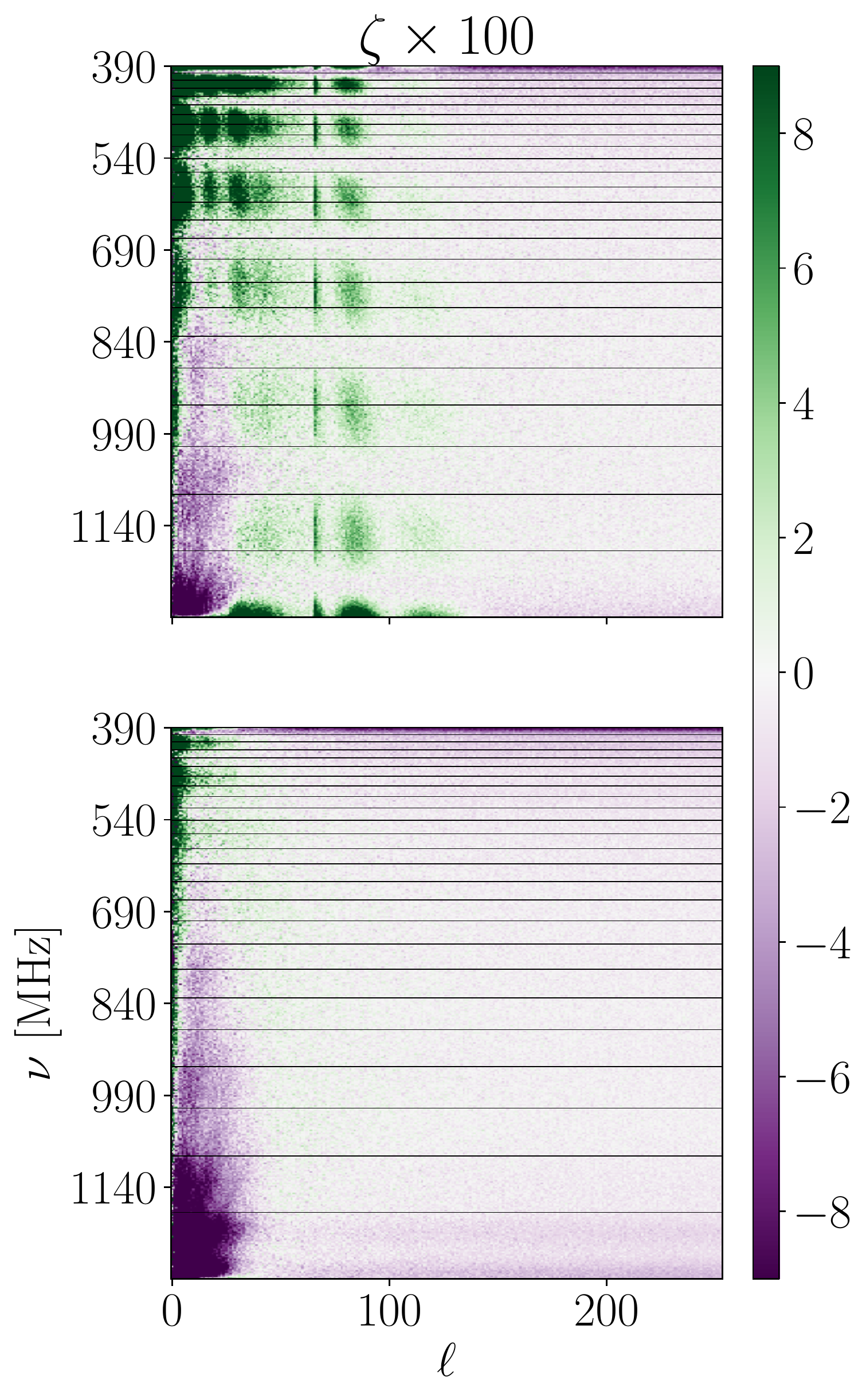}
      \caption{We demonstrate the effectiveness of foreground cleaning with 7 (top) and 9 degrees of freedom (bottom). The relative difference $\zeta$ is defined in Eq. \ref{eq:zeta}. For our purposes (measuring large scales), cleaning with 7 degrees of freedom is clearly not sufficient as it leaves residuals on scales up to $\ell \lesssim 100$. Therefore the choice of $N_{\rm FG}=9$ is adopted throughout this work unless otherwise stated. The horizontal lines indicate the frequency binning used in this analysis, corresponding to a fixed width in redshift of $\Delta z = 0.1$.} \label{fig:fg_rm}
    \end{figure}
    \begin{figure}
      \includegraphics[width=1.0\columnwidth]{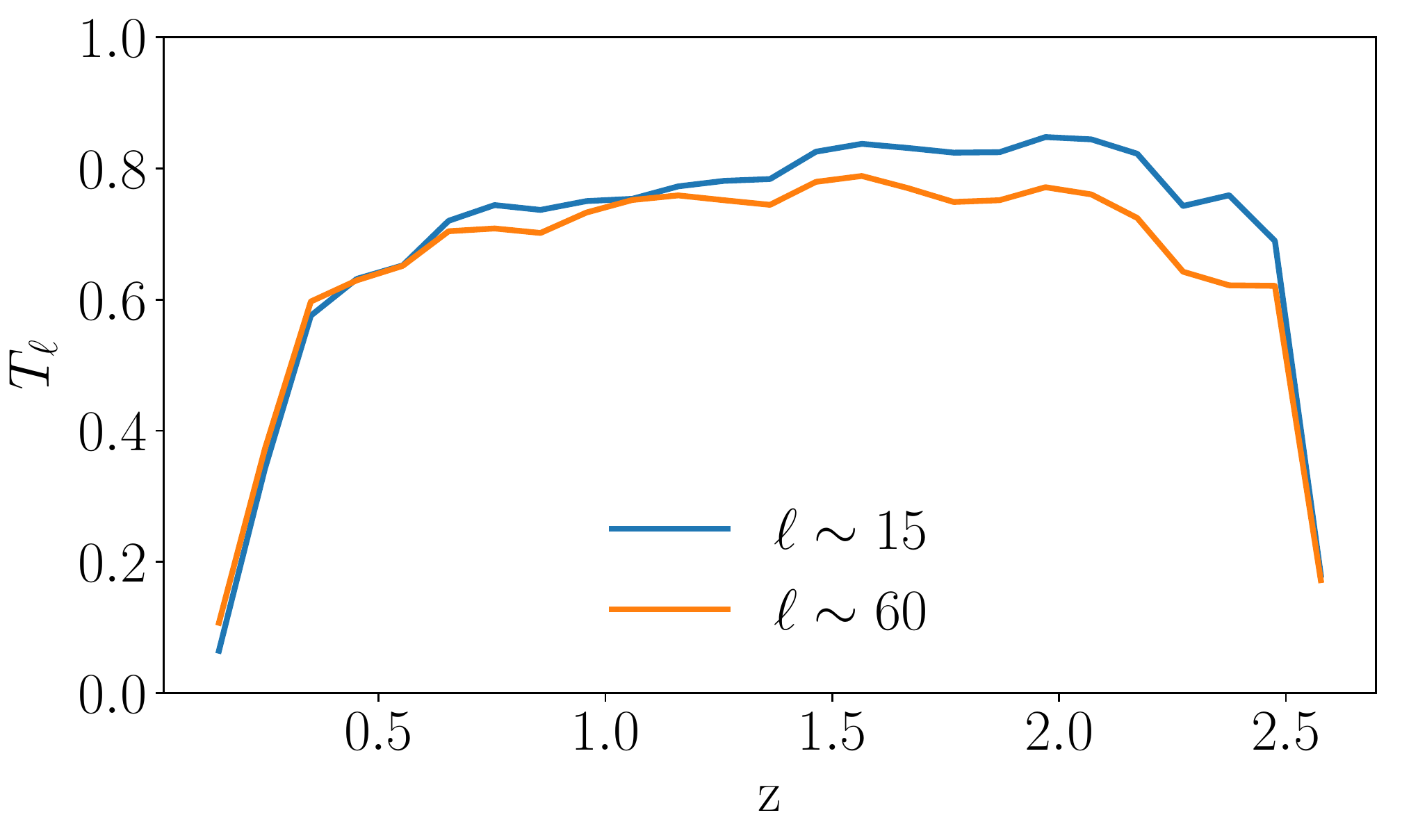}
      \caption{The foreground removal transfer function $T_\ell(z)$ for $9\leq \ell \leq 21$ (blue) and $54 \leq \ell \leq 66$ (orange). The transfer function shows only a mild scale dependence, but drops significantly at the edges of the redshift range, where foreground removal is less efficient (see \citep{doi:10.1093/mnras/stu2474}).} \label{fig:bandpass}
    \end{figure}
    To remove the foregrounds from our simulations, we use the Principal Component Analysis method (PCA), as described in \cite{doi:10.1093/mnras/stu2474}. In short, the method is based on de-projecting the principal eigenmodes of the frequency-frequency covariance matrix estimated from the data, under the assumption that those modes are the ones most contaminated by foregrounds. The level of conservativeness in the foreground removal stage can be parametrized by the number of de-projected modes, which we will refer to as the number of foreground degrees of freedom $N_{\rm FG}$.
    
    In order to estimate the number of foreground degrees of freedom that must be de-projected in our simulations, we ran the foreground removal algorithm on all of them for different values of $N_{\rm FG}$. For each value, we use, as a diagnostic for foreground contamination, the relative systematic deviation in the angular power spectrum as a function of frequency and angular scale, defined as
    \begin{equation}\label{eq:zeta}
      \zeta_\ell(\nu)=\left\langle\frac{C_\ell^{\rm clean}(\nu)}{C_\ell^{\rm free}(\nu)}-1\right\rangle.
    \end{equation}
    Here $C^{\rm free}_\ell$ and $C_\ell^{\rm clean}$ are the power spectra for foreground-free simulations  and for simulations in which $N_{\rm FG}$ foreground modes have been subtracted respectively. The optimal $N_{\rm FG}$ was then determined as the minimum value that achieves an acceptable degree of foreground removal over a large fraction of the $\ell-\nu$ plane. This quantity is shown in Fig. \ref{fig:fg_rm} for the cases $N_{\rm FG}=7$ and $N_{\rm FG}=9$. Green colours represent a higher power spectrum with respect to the true one, and are a sign of foreground contamination, while purple areas represent lower power spectrum amplitudes and denote a loss of signal-dominated modes caused by over-fitting. As mentioned in Section \ref{ssec:method.fgs}, the latter effect can be corrected analytically once the foreground removal transformation has been established (e.g. through the transfer function $T_\ell$), and therefore we seek to minimize foreground contamination. In view of the results shown in this figure, we chose to use $N_{\rm FG}=9$ as our fiducial value. The transfer function associated with this choice of $N_{\rm FG}$, as defined in Section \ref{ssec:method.fgs}, is shown in Figure \ref{fig:bandpass} for all different redshift bins as a function of scale.

  \subsection{Results in the presence of foregrounds}\label{ssec:res.wfg}
    As described in Section \ref{ssec:method.fgs}, the effect of foregrounds is two-fold:
    \begin{enumerate}
      \item Foreground contamination in the auto-correlation will lead to a bias in $\auto$ that can be statistically significant;
      \item Foreground removal will erase some of the long-wavelength modes in the signal. This reduces the number of common modes between the foreground-cleaned intensity maps and the galaxy distribution, thereby degrading the performance of the multi-tracer technique.
    \end{enumerate}
    We first quantify these two effects and then elaborate on their root causes and possible ways around them.
    
    \subsubsection{Sensitivity and bias}\label{sssec:res.wfg.sb2n}
      \begin{figure}
        \includegraphics[width=1.0\columnwidth]{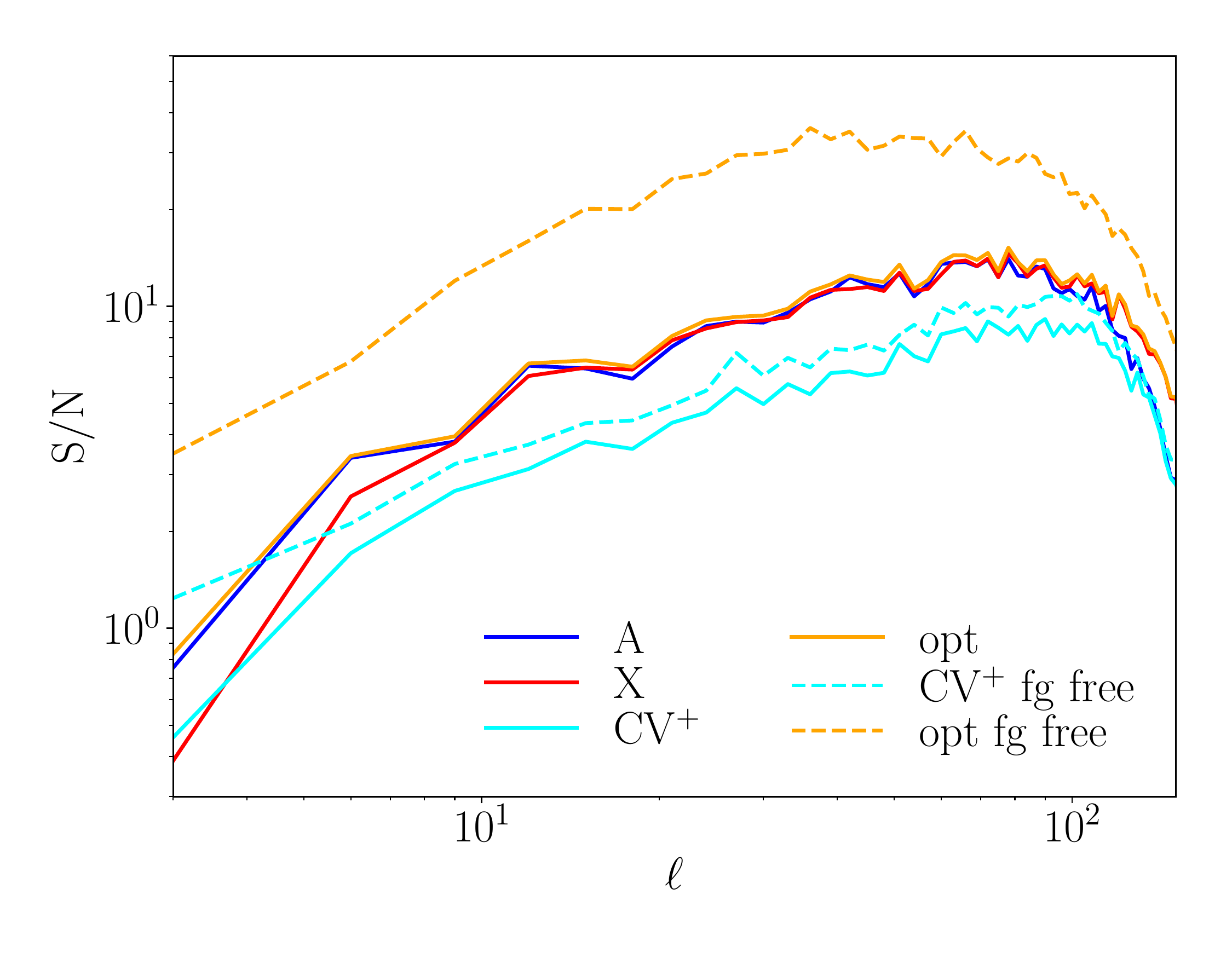}
        \caption{The signal-to-noise ratio for all estimators in the full analysis plus $\opt$ and $\CV$ in the foreground-free case, as references. The inclusion of foregrounds in the analysis introduces a significant degradation in sensitivity, and only a slight improvement (a factor $\sim1.5$) over the foreground-free cosmic-variance limit (dashed cyan line) is possible. }
        \label{fig:estfull}
      \end{figure}
      \begin{figure}
        \includegraphics[width=1.0\columnwidth]{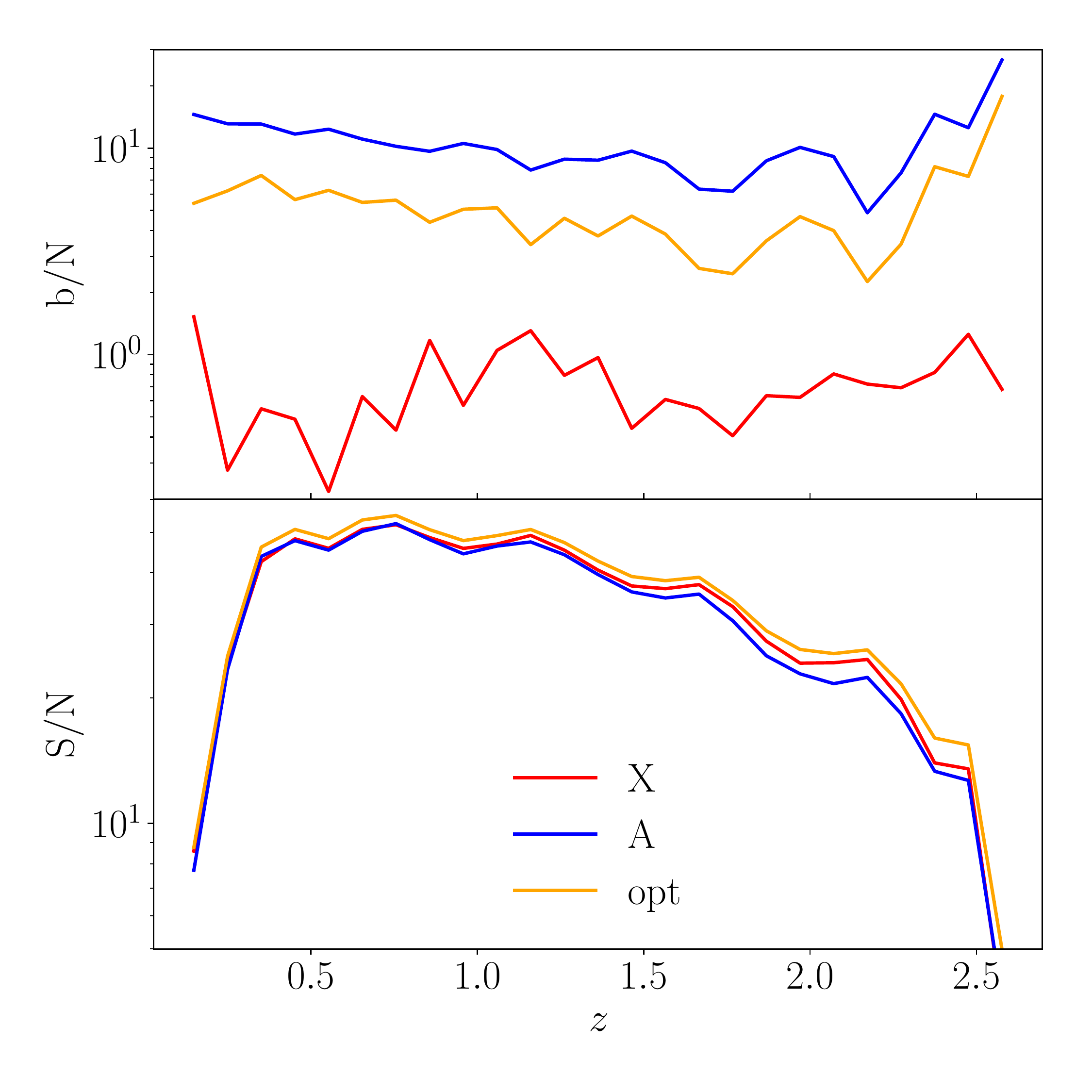}
        \caption{The signal-to-noise (bottom) and bias-to-noise ratio (top) as a function of redshift, for $\auto$ (blue), $\cross$ (red) and $\opt$ (yellow) in the full analysis. The bias for $\opt$ and $\auto$ increases for low and high redshifts, because foreground cleaning is less effective there (see also figures \ref{fig:fg_rm} and \ref{fig:bandpass}). The bias in $\cross$ is compatible with 1$\sigma$ fluctuations, thanks to lack of foreground residuals in the ${\rm HI-galaxy}$ cross-correlation. Foreground cleaning still introduces a random error in all estimators, which is highest at the upper and lower ends of the frequency range, similar for all estimators.}
        \label{fig:bias}
      \end{figure}
      \begin{figure*}
        \includegraphics[width=1.0\columnwidth]{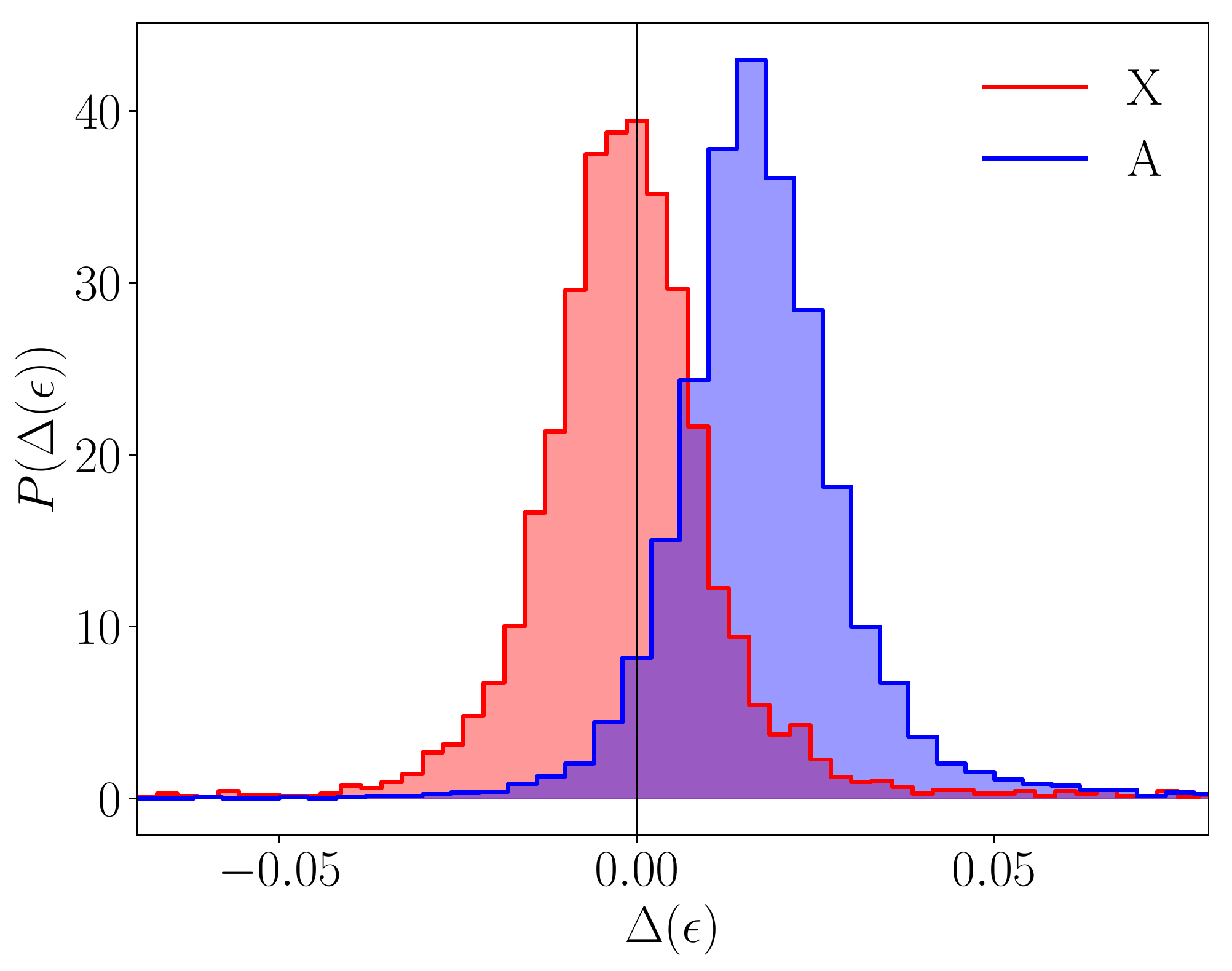}
        \includegraphics[width=1.0\columnwidth]{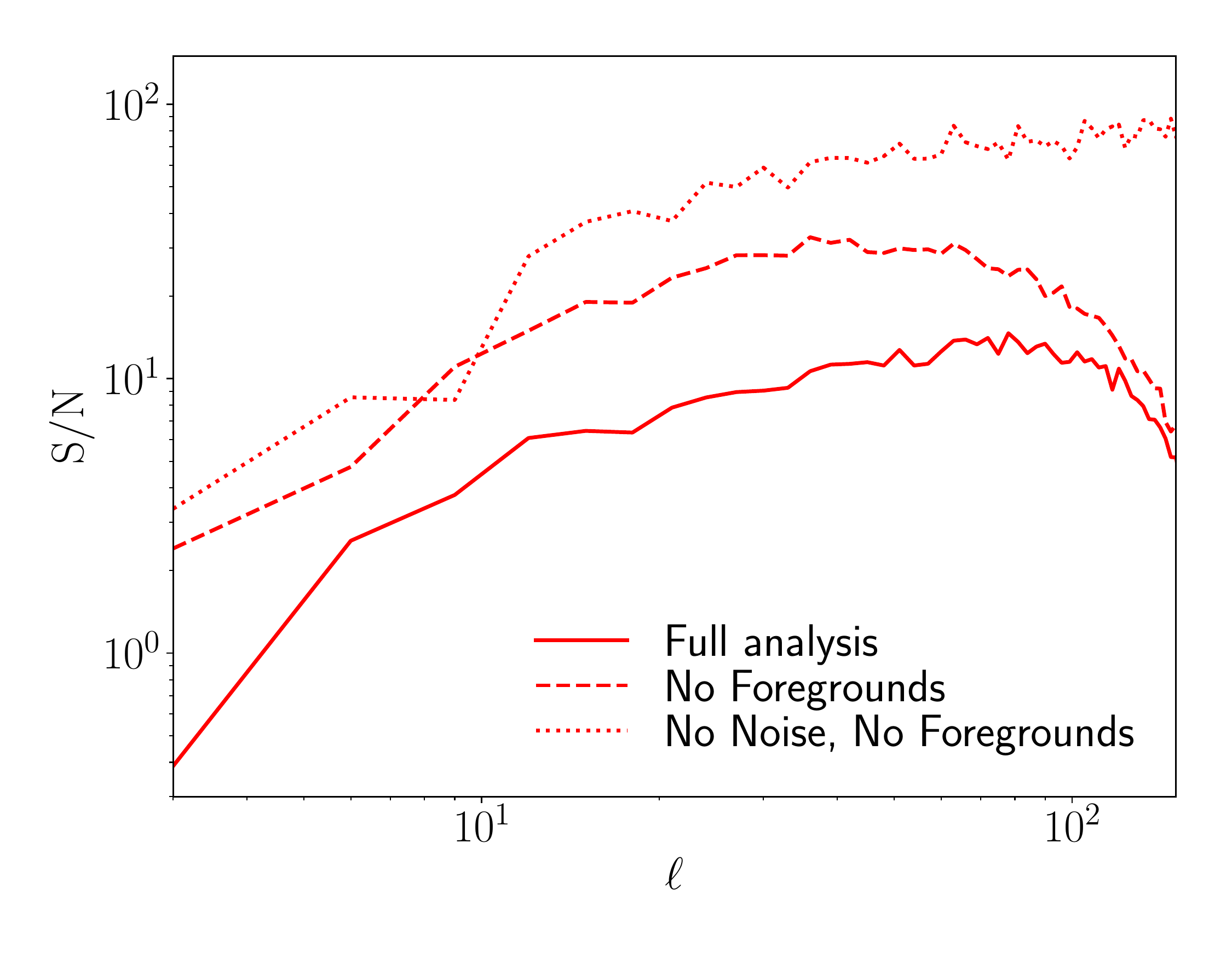}
        \caption{{\sl Left panel:} Distribution of $\Delta(\est) = \est - \est^\mathrm{true}$ for $\auto$ (blue) and $\cross$ (red) in the full analysis on the range $0 \leq \ell \leq 81$. The systematic error in $\auto$ is due to the effects of foreground cleaning combined with the low radial resolution of the galaxy maps. {\sl Right panel:} The signal-to-noise ratio of $\cross$ for the full analysis (solid line), without foregrounds (dashed line) and without foregrounds or noise (dotted line). At large scales up to $\ell \lesssim 80$ foregrounds are the dominant source of uncertainty for $\cross$.}
        \label{fig:dist}
      \end{figure*}
      The impact of the loss of long-wavelength modes in the method's sensitivity can be studied through the signal-to-noise ratio defined in Section \ref{ssec:res.nofg}. The results are shown in Fig. \ref{fig:estfull} as solid lines for $\auto$, $\cross$ and $\opt$ as a function of scale for a redshift bin at $z=1$. The figure also shows the results for the cosmic-variance dominated estimator $\CV$ described in Section \ref{ssec:res.nofg} as a solid cyan line. When comparing with the CV limit in the presence of foregrounds we observe that all estimators are able to improve upon $\CV$, although now only by a factor of $\sim2$. However, when comparing with the full constraining power in the absence of foregrounds, shown as a dashed orange line for $\opt$ and as a dashed cyan line for $\CV$ in the same figure, we observe a significant loss in $S/N$ and that the impact of foregrounds prevents the estimators from producing a significant improvement in sensitivity with respect to an analysis without CV-cancellation (as would be the case of a single-tracer galaxy survey). We explore this effect in more detail below.
    
      To explore the first effect described at the beginning of this section (the foreground bias), we start by defining the ``bias-to-noise'' ratio for a given estimator as
      \begin{equation}
       \left(\frac{b}{N}\right)_\ell\equiv\frac{\langle\est_\ell\rangle-\est^{\rm true}_\ell}{\sigma_\ell},
      \end{equation}
      where $\est_\ell^{\rm true}$ and $\sigma_\ell$ are defined in Section \ref{ssec:res.nofg}. We compute this quantity for all redshift bins and multipoles, and then estimate a scale-averaged ${\rm b}/{\rm N}$ 
      as a quadrature sum of the $\ell$-dependent ratio
      \begin{equation}
        \left(\frac{b}{N}\right)(z)=\sqrt{\sum_{\ell = 0}^{81} \left(\frac{b}{N}\right)^2_\ell(z)},
      \end{equation}
      where the sum is taken over the range of relevant multipoles. This quantity is shown in the upper panel of Fig. \ref{fig:bias} as a function of redshift for the three estimators considered here. While the bias of the cross-correlation-based estimator $\cross$ is compatible with $\sim1\sigma$ fluctuations, the use of auto-correlations through either $\auto$ or $\opt$ produces noticeable biases of up to $10\sigma$, caused by foreground contamination. The lower panel of the same figure shows the integrated $S/N$ ratio (estimated as a quadrature sum over power spectrum multipoles), and reinforces our conclusion that all estimators achieve similar sensitivities, and therefore we do not incur in any significant loss by dropping all auto-correlation information and using only cross-correlations for which foregrounds do not induce any bias.
      
      Finally, we summarize the main findings of this section in Fig. \ref{fig:dist}. The left panel shows the distribution of $\auto-\est^{\rm true}$ and $\cross-\est^{\rm true}$ across all simulations and $\ell$ values for a bin at $z\sim0.8$. The distributions are close to Gaussian, and the $\auto$ shows a clear foreground bias. The right panel shows the degradation in sensitivity caused by instrumental noise (dotted line to dashed line) and by the partial removal of signal due to foregrounds (dashed line to solid line).
      
    \subsubsection{The effects of foregrounds}\label{sssec:res.wfg.more}
      \begin{figure}
        \includegraphics[width=1.0\columnwidth]{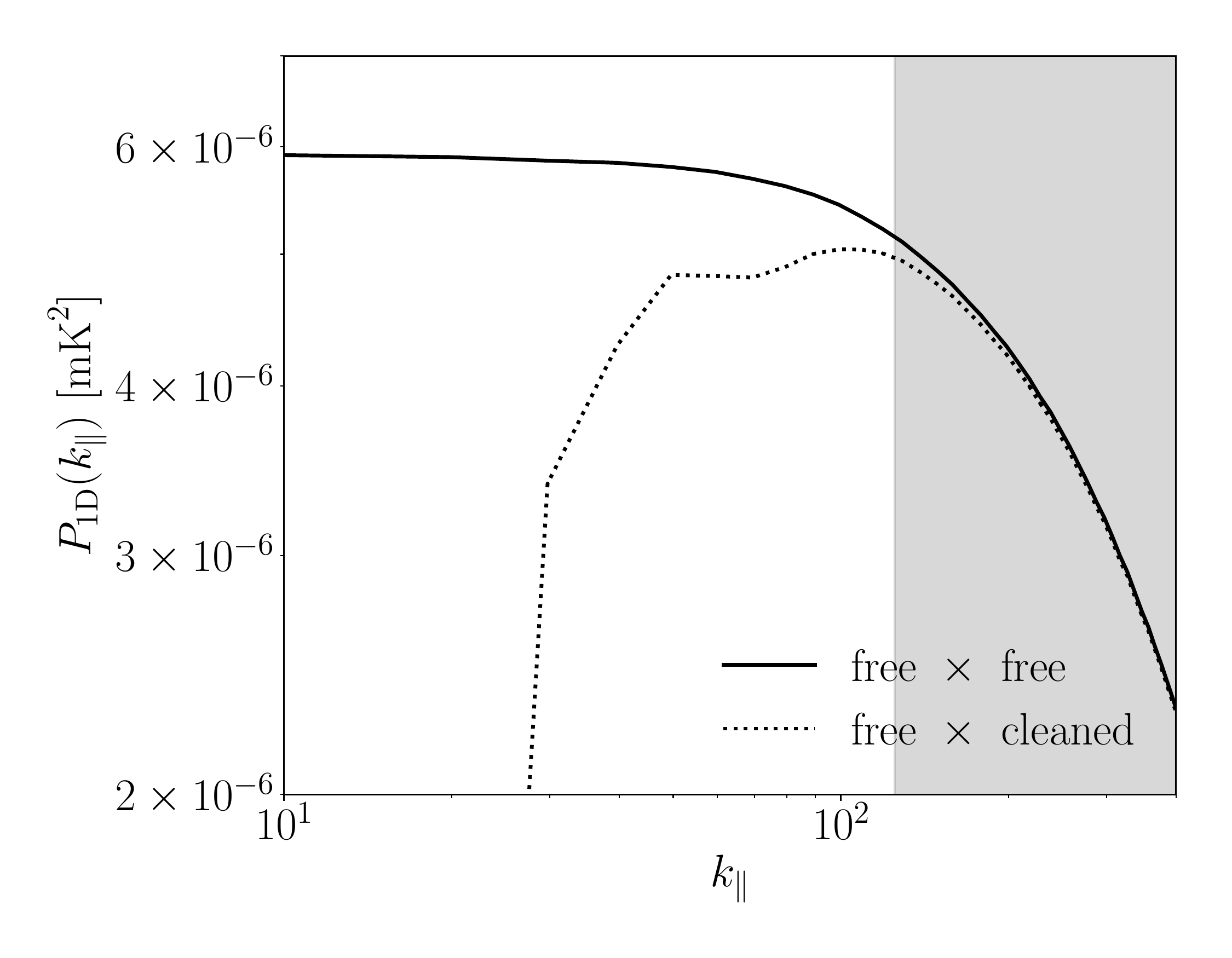}
        \caption{Radial HI power spectrum, averaged over $200$ simulations, showing the auto-correlation of foreground free-maps (solid line) and the cross-correlation of foreground-free and foreground-cleaned maps (dotted line). Note that we use a non-standard dimensionless radial coordinate $\nu/\nu_{21}$ (see Eq. \ref{eq:fftr}), and therefore the wave number $k_\parallel$ is also dimensionless. The loss of long-wavelength radial modes is apparent in the drop of the dotted line for $k_\mathrm{\parallel} \lesssim 100$. The grey shaded area indicates the smoothing scale due to the redshift bin width of $\Delta z = 0.1$, associated with the LSST photo-$z$ uncertainty. Unfortunately this is where foreground cleaning works best and the solid and dotted lines agree, limiting the scale overlap between both types of observations.}
        \label{fig:pkrad}
      \end{figure}
      \begin{figure}
        \includegraphics[width=1.0\columnwidth]{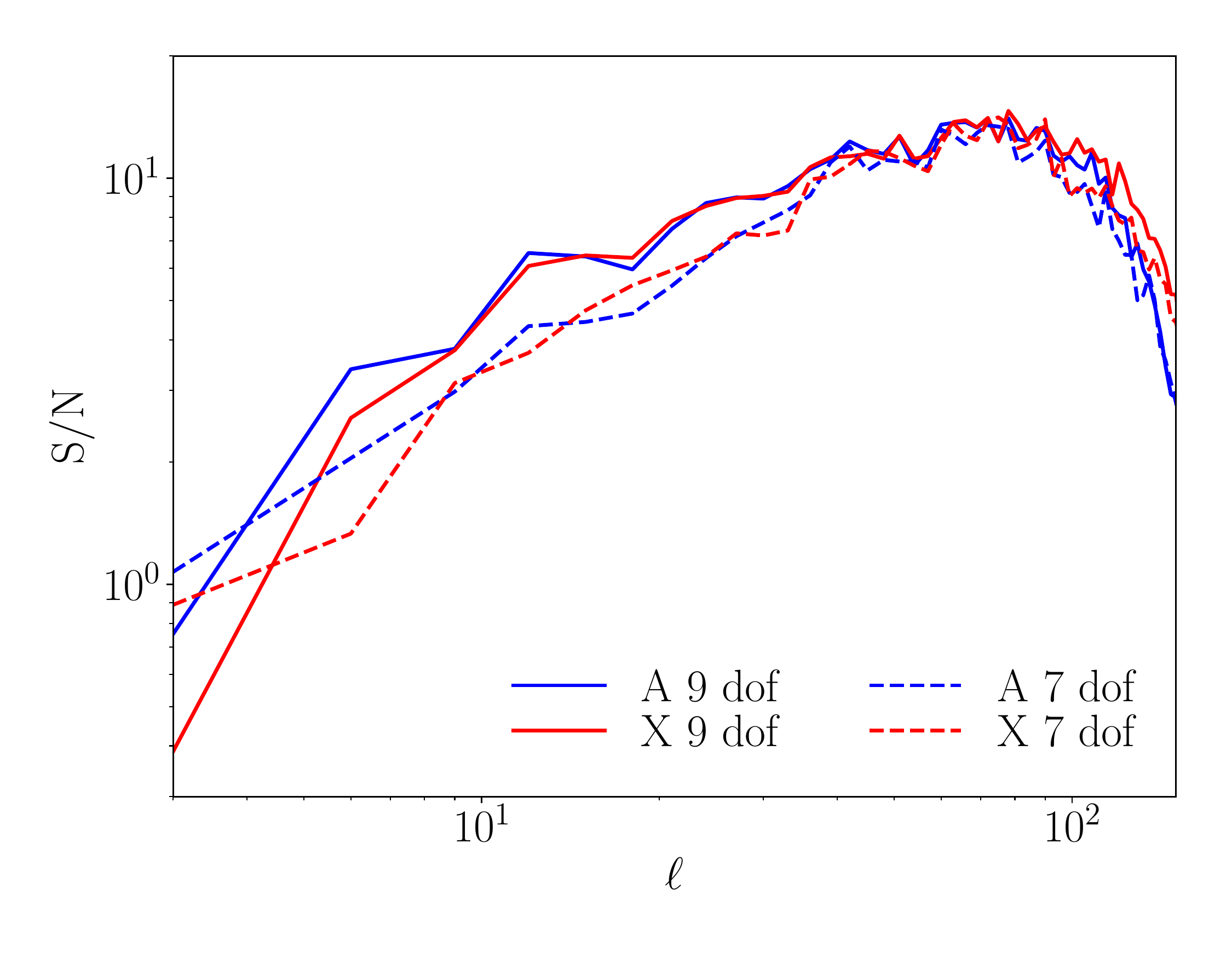}
        \caption{The signal-to-noise ratio for $\cross$ (red) and $\auto$ (blue) with noise and foregrounds, comparing results obtained from cleaning with 9 (solid lines) and 7 foreground degrees of freedom (dashed lines). The contribution of foreground residuals to the estimator noise outweighs the potential improvement in sensitivity due to the milder subtraction of long-wavelength modes, and the $N_{\rm FG}=7$ case yields a poorer results than our fiducial choice of $N_{\rm FG}=9$.}
        \label{fig:9VS7}
      \end{figure}
      We have carried out a number of tests to further understand the effects of foregrounds on multi-tracer analyses, and to explore different avenues to mitigate these effects.
      
      As we have seen, the cross-correlation estimator $\cross$ is immune to foreground bias and its use does not incur in any significant penalty in terms of sensitivity. Therefore, the main impact of foregrounds in 21cm observations is the loss of long radial wavelength modes present in the galaxy distribution. To quantify this effect we have studied the radial 1D power spectrum $P_{\rm 1D}(k_\parallel)$, defined as the variance of the line-of-sight Fourier coefficients of our 21cm maps. In practice we estimate this observable, as outlined in \cite{2017MNRAS.466.2736V}, by computing the Fourier transform for a given pixel across all frequencies:
      \begin{equation}\label{eq:fftr}
        \Delta T(k_\parallel,\nv)=\int \frac{d\nu}{\nu_{21}\sqrt{2\pi}}\exp\left[i\frac{\nu k_\parallel}{\nu_{21}}\right]\Delta T(\nu,\nv).
      \end{equation}
      Note that we use $\nu/\nu_{21}$, as a radial coordinate, where $\nu_{21}=1420\,{\rm MHz}$ is the frequency of the 21cm line, and therefore the radial wavenumber $k_\parallel$ is dimensionless\footnote{In practice the Fourier transform is computed as a discrete Fourier transform \citep{FFTW05}.}. The 1D power spectrum is then computed as the covariance between two fields $\Delta T_1$ and $\Delta T_2$:
      \begin{equation}
        P_{1D}(k_\parallel)=\frac{\Delta\nu}{\nu_{21}2\pi}\left\langle{\rm Re}\left[\Delta T_1(k_\parallel)\Delta T^*_2(k_\parallel)\right]\right\rangle,
      \end{equation}
      where the average is taken across all unmasked pixels and all simulations.
      
      Figure \ref{fig:pkrad} shows two 1D power spectra, computed from the auto-correlation of the foreground-free simulations (solid line) and from the cross-correlation of the foreground-clean and foreground-free simulations (dotted line). Although both power spectra match on small scales ($k\gtrsim200$), the loss of long-wavelength radial modes becomes apparent on larger scales, where the amplitude of the cross-correlation becomes significantly smaller than the foreground-free power spectrum. On the other hand, the radial smearing effect of photometric redshifts will erase all structure on scales smaller than the photo-$z$ error $\sigma_z$. Since $\nu/\nu_{21}=(1+z)^{-1}$, we can associate $\sigma_z$ with a threshold wavenumber $k_{\rm ph}\equiv\pi(1+z)^2/\sigma_z$. At $z\sim1$ and assuming $\sigma_z=0.03\,(1+z)$, we obtain $k_{\rm ph}\sim200$, which coincides with the scale at which the mode loss to foregrounds becomes noticeable. The bin width $\Delta z=0.1$ would correspond to a scale $k_\parallel\sim125$, and so effectively all the modes within the shaded region of Fig. \ref{fig:pkrad} are erased in the data, due to the top-hat smoothing. The range of radial scales over which a significant overlap between 21cm observations and an LSST-like galaxy sample can be found becomes significantly reduced, which has a negative impact on the cosmic variance cancellation of the estimators studied here.
      \begin{figure}
        \includegraphics[width=1.0\columnwidth]{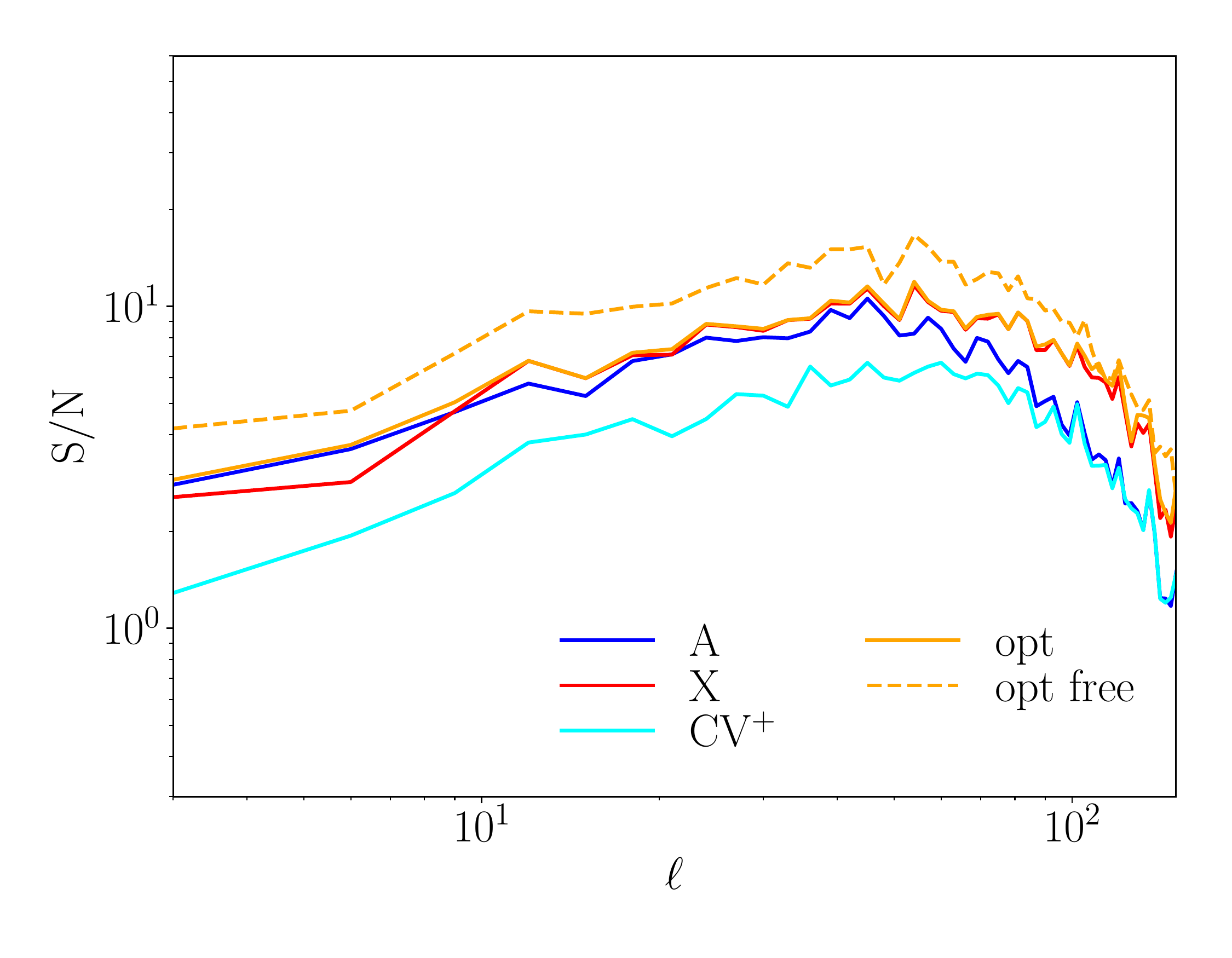}
        \includegraphics[width=1.0\columnwidth]{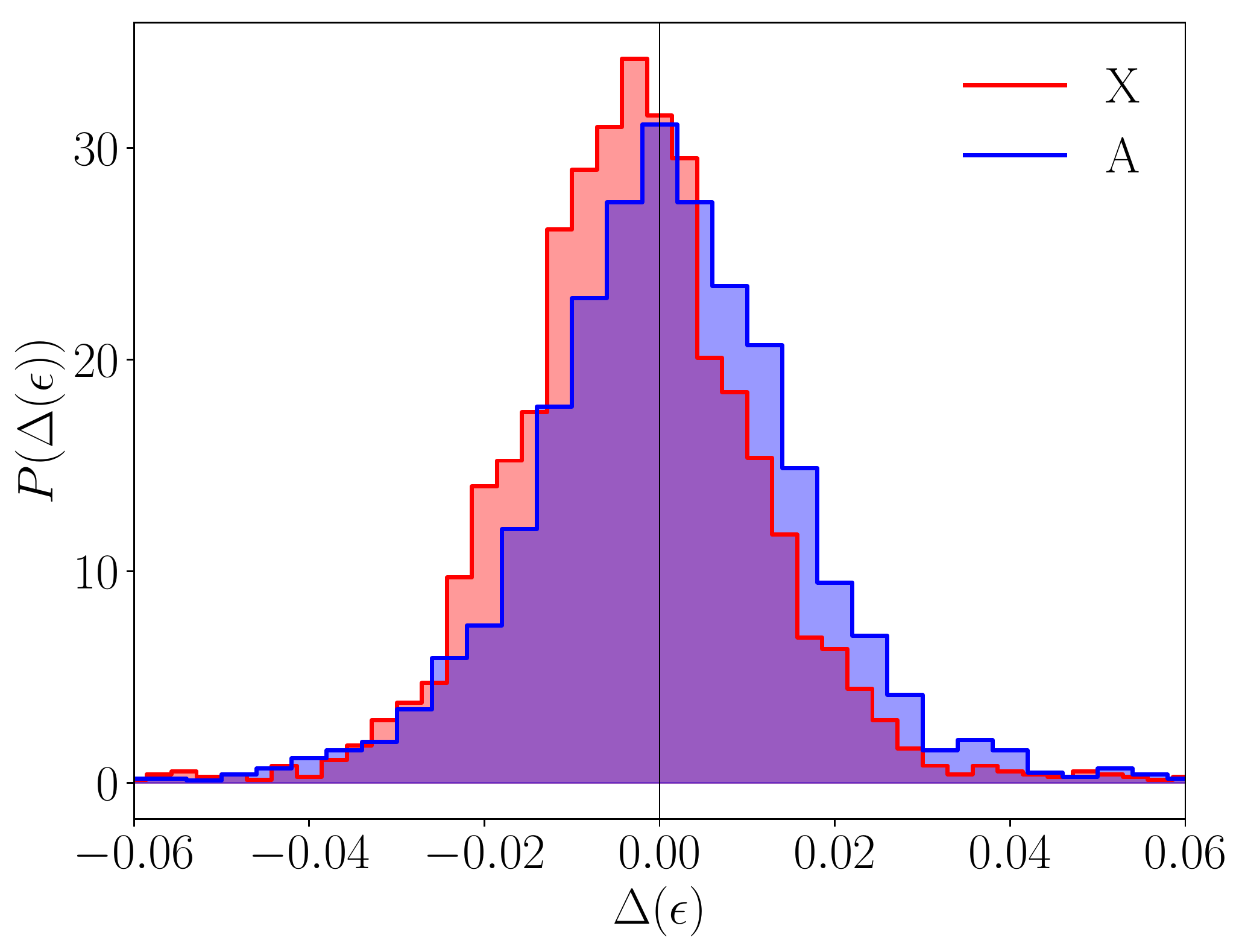}
        \caption{{\sl Top panel:} Similarly to fig. \ref{fig:estfull}, but working in thin bins of $\Delta \nu = 1$ MHz for both the galaxy and HI maps. We show the signal-to-noise ratio for all estimators in the full analysis, and $\opt$ in the foreground free case. The latter is worse due to the smaller information content in the much thinner bin, but notably all solid lines actually outperform the respective results obtained from thick bins. This is due to the much bigger overlap of radial modes in the galaxy and HI power spectra when precise galaxy redshifts are assumed. Note that the full constraining power in this case would be realized by combining the information from the much larger number of redshift bins. {\sl Bottom panel:} same as fig. \ref{fig:dist}, but also for the thin frequency bin. While the shape and size of the distributions of $\cross$ (red) and $\auto$ (blue) hardly change, the bias in $\auto$ becomes negligible, due to the larger relative number of overlapping radial modes between the 21cm and galaxy data.}
        \label{fig:s2n_nomerge}
      \end{figure}
      
      To circumvent this problem we have explored a few alternative avenues:
      \begin{itemize}
        \item {\bf Foreground degrees of freedom.} To reduce the number of modes lost to foreground removal it is worth exploring the possibility of subtracting a smaller number of degrees of freedom at that stage. As discussed in Section \ref{ssec:res.fgrm}, this will produce significant foreground residuals that will bias the auto-correlation but, since $\cross$ is immune to this bias, its sensitivity might benefit significantly from the presence of additional signal modes. However, although the foreground residuals will not contribute to the bias of $\cross$, they will also provide a contribution to its variance, and therefore there will be a balance between the preservation of long-wavelength modes and the contribution of foreground residuals to the noise.
        
        Figure \ref{fig:9VS7} shows the $S/N$ ratio of $\auto$ (blue) and $\cross$ (red) for the fiducial case, in which $N_{\rm FG}=9$ foreground degrees are subtracted (solid lines) and for an alternative scenario with $N_{\rm FG}=7$ (dashed lines). No significant improvement is obtained in both cases, and in fact we observe a reduction in sensitivity on large scales. Therefore, at least for this setup, the contribution of foreground residuals to the estimator variance outweighs the impact of the additional signal degrees of freedom allowed by a more lenient removal stage. More efficient foreground removal methods preserving more information from the signal while at the same time removing all residuals on large scales could potentially improve this result.
        
        \item {\bf Thinner redshift bins.} The large photo-$z$ uncertainties that can realistically be achieved by an experiment like LSST will make the use of redshift bins smaller than $\Delta z\sim0.1$ pointless. This is due to the strong correlations between narrower bins induced by the photo-$z$ scatter. Nonetheless, it is worth exploring the possible cosmic-variance cancellation gains if a better redshift precision were available. To do so, we have repeated our analysis making use of redshift bins with width $\Delta z=0.02$.
        
        The results of this exercise are shown in Fig. \ref{fig:s2n_nomerge} for a bin centered around $z\sim0.8$. The the upper panel corresponds to the signal-to-noise as a function of scale, and shows an improvement of a factor $\sim2$ with respect to the cosmic-variance-limited case for all estimators. Although this is comparable with the results shown for individual bins with $\Delta z=0.1$, the number of uncorrelated bins available in this case is 5 times larger, and therefore the total signal-to-noise increases significantly. The lower panel of the same figure shows the distributions of $\auto$ and $\cross$ for the same redshift bin across all simulations and values of $\ell$. The significant bias in $\auto$ observed in Fig. \ref{fig:dist} is now gone, owing to the relative increase in radial modes on which the 21cm signal dominates over foreground residuals.
        
        As we emphasized above, although higher redshift resolution improves the performance of multi-tracer methods for 21cm intensity mapping, photometric redshift surveys are unlikely to achieve the required redshift accuracy. On the other hand, although spectroscopic surveys can easily reach that level of radial resolution, they can only do so for a substantially smaller number of objects, and the larger shot noise will inevitably affect the performance of the multi-tracer technique. The most promising option is perhaps intensity mapping of other emission lines (e.g. the CO line \citep{2018MNRAS.475.1477P}), as long as the instrumental noise can be reduced sufficiently.
        
        \item {\bf Matching scales.} Finally, another possibility would be to subject the galaxy overdensity data to the same linear transformation that downweights the long wavelength modes in the 21cm maps. If this can be done with sufficient accuracy, the resulting auto and cross-power spectra should manifest the same fluctuations around the mean on a realization-by-realization basis, and a higher degree of CV cancellation could be expected from the estimators used here, based on ratios of those quantities.
        
        Note that in a more optimal analysis, where the full data from the 21cm maps and the galaxy overdensity are used, including all the signal-dominated radial and angular modes (instead of just the power spectrum ratios of matching redshift bins), this is unlikely to provide any advantage over preserving all of the modes in the latter probe. A likelihood evaluation of the full data would automatically produce the cancellation of cosmic variance on all common modes, and would use the additional galaxy long-wavelength modes to increase the final constraints further.
      \end{itemize}
      
      Thus, to summarize: although multi-tracer methods applied to 21cm data in cross-correlation with photometric redshift surveys do improve the constraining power beyond the cosmic variance limit, this improvement is strongly hampered by the loss of long-wavelength modes, common to both datasets, due to the presence of foreground contamination and low $z$ resolution. Multi-tracer analyses using 21cm observations are therefore more likely to achieve a better performance when combined with other intensity mapping data or spectroscopic surveys, assuming the noise amplitude of the latter (instrumental or shot-noise) can be reduced sufficiently. 

\section{Discussion}\label{sec:discussion}
  \begin{table}
    \begin{center}
      \begin{tabular}{l|rrrr}
        \hline
        Case & $\auto$ &  $\cross$ & $\CV$ & $\opt$\\
        \hline
        No noise, no FG & 1291 & 1292 & 192 & 1306 \\
        No FG           &  495 &  502 & 154 &  509 \\
        No noise        &  299 &  298 & 155 &  312 \\
        Full analysis   &  178 &  183 & 120 &  192 \\
        \hline
      \end{tabular}
      \caption{Signal-to-noise from combining all redshift bins for all estimators and all modelling scenarios of this work. Here using 9 degrees of freedom for the foreground cleaning and a redshift bin width of $\Delta z = 0.1$. While $\auto$ uses the HI and galaxy auto-correlations, $\cross$ uses the HI-g cross-correlation and g-g auto-correlation and $\opt$ is the inverse variance-weighted sum of both (eqs. \ref{eq:est_e1} - \ref{eq:est_opt}). The estimator $\CV$ on the other hand uses auto-correlations with different DM realizations for the galaxy and HI populations and shows the constraints achievable in the absence of multi-tracer cosmic variance cancellation.}
      \label{tab:S2N}
    \end{center}
  \end{table}
    
  21cm intensity mapping and photometric redshift surveys are two promising techniques to study the three-dimensional distribution of matter in the Universe on large scales. A number of cosmological observables, such as the level of primordial non-Gaussianity, benefit from the combined analysis of multiple proxies of the same density inhomogeneities in what is known as the ``multi-tracer'' technique. In this paper we have explored the feasibility of multi-tracer analyses that exploit the combination of the two aforementioned probes, for the particular case of 21cm observations to be carried out by an SKA-like instrument and an LSST-like galaxy sample. For concreteness, we have focused our analysis on two estimators of the bias ratio for both samples, $\auto$ and $\cross$, described in Section \ref{ssec:theory.estimators}. Since these estimators make use of the 21cm auto-correlation and its cross-correlation with galaxies respectively, they allow us to explore both the bias induced on $\auto$ by the presence of foreground residuals, and the potential loss of information associated with dropping auto-correlation information ($\cross$). For completeness, we also consider an optimal inverse-variance combination of both estimators, $\opt$, that uses all the data available.
    
  In the absence of foregrounds, we show that both $\auto$ and $\cross$ are able to achieve similar sensitivities, with little improvement when using $\opt$ due to the tight correlation between both estimators. When compared with the a cosmic-variance contaminated measurement of the same bias ratio, we show that these estimators are able to improve the signal-to-noise by a factor of $\sim4$-$5$, even when compared to the cosmic-variance-contaminated, noise-free case, showcasing the tremendous potential gains of the multi-tracer technique.
    
  The impact of the presence of foregrounds in the 21cm data is twofold. On the one hand, residuals after foreground removal produce an offset in the \textsc{HI} auto-correlation which biases both $\auto$ and $\cross$ at high significance. We show however, that $\cross$ is immune to this bias, while preserving the same statistical power as the two other estimators. On the other hand, foreground removal is based on the separation of foregrounds and cosmological signal through their different spectral behaviour, effectively down-weighting the radial long-wavelength modes where foregrounds dominate. Since photometric redshifts effectively erase all structure along the line of sight on all but the largest scales, the overlap between SKA and LSST in the $k_\parallel$-$k_\perp$ plane reduces significantly, partially spoiling the cosmic variance cancellation. We show that, in this case, the sensitivity of all estimators drops by more than a factor of $\sim2$, and that the improvement in signal-to-noise ratio with respect to a cosmic-variance-limited measurement made in the same circumstances is now only a factor $\sim2$. This drops to a smaller $\sim50\%$ improvement when we compare either estimator with the cosmic-variance-limited measurement without foregrounds. These results are summarized in Table \ref{tab:S2N}, which shows the cumulative signal-to-noise (quadrature-summed over all multipoles and redshift bins) for the three estimators as well as the CV limit in different scenarios regarding the presence of noise and foregrounds.
    
  We have also explored two possible ways to overcome this problem. First, a less aggressive foreground removal that leaves a larger fraction of foreground residuals in the maps, would also leave a larger number of long-wavelength modes untouched, increasing the scale overlap between LSST and SKA. In practice, however, we have seen that the contribution of the foreground residuals to the estimator uncertainties in fact decrease the total SNR when a smaller number of foreground degrees of freedom are subtracted. Another way to increase the scale overlap between both experiments would be to reduce the size of the redshift bins used in the analysis. Although this is not a real possibility for photometric surveys, since structures can never be resolved on scales smaller than the photo-$z$ uncertainty, this case allows us to explore other possible synergies with either spectroscopic surveys or intensity mapping observations of other emission lines. Our results show that in this case the gain in sensitivity associated with the multi-tracer technique is likely restored, with the added advantage that the foreground bias is also reduced due to the larger fraction of signal-dominated modes. Since the shot noise associated with currently planned wide-area spectroscopic surveys is likely to be too large for the multi-tracer technique to be effective, we argue that the most promising way forward for these types of analysis may be the combination of intensity mapping observations for different emission lines.
    
  In a follow up work we plan to study these new avenues in more detail, first by considering constraints on $\fnl$ directly, and including estimators that can deal naturally with the mismatch in the modes that are removing in different surveys does allowing a more perfect cancellation. We will also consider other foreground cleaning methods that might be less aggressive on cleaning this large scales and new tracers with higher redshift resolution.

\section*{Acknowledgments}
  We would like to thank Pedro Ferreira for useful comments and discussions. AW, JF and MGS acknowledge support from the South African Square Kilometre Array Project and National Research Foundation. AW was also supported by the Newton Fund PhD Partnerhship Scheme, NRF grant 98555 and UK grant ES/N013956/1, part of the UK’s official development assistance (ODA). DA acknowledges support from the Beecroft Trust and from the Science and Technology Facilities Council (STFC) through an Ernest Rutherford Fellowship, grant reference ST/P004474/1. We acknowledge the support of the Centre for High Performance Computing, South Africa, under the project ASTR0945.

\bibliographystyle{mnras}
\bibliography{mtdraft}

\end{document}